\documentclass[a4paper,10pt]{article}
\pdfoutput=1 

\usepackage{jheppub} 

\usepackage{float}
\usepackage{graphicx}
\usepackage{subcaption}
\usepackage{bbm}
\usepackage{siunitx}
\usepackage{listings}
\usepackage{centernot}

\usepackage{graphicx,epsf}
\usepackage{amsmath,amsfonts,amssymb,amsbsy}
\usepackage{mathtools}
\usepackage{physics}

\usepackage[dvipsnames]{xcolor}

\usepackage{hyperref}
\usepackage{xcolor}

\usepackage{multirow}
\usepackage{stackrel}
\usepackage{tikz}
\usetikzlibrary{calc}

\usepackage{ifdraft}
\usepackage[obeyFinal]{easy-todo}

\usepackage{placeins}
\usepackage{slashed} 

\usepackage{booktabs}
\usepackage{adjustbox}

\usepackage[utf8]{inputenc}

\usepackage{graphicx}
\usepackage{pdfpages}
\usepackage{makecell}

\usepackage{cleveref}
\usepackage{pdflscape}

\usepackage{pifont}

\usepackage{enumitem}

\newcommand{\be}{\begin{equation}}
\newcommand{\ee}{\end{equation}}

\newcommand{\bea}{\begin{eqnarray}}
\newcommand{\eea}{\end{eqnarray}}

\def\spa#1.#2{\left\langle#1\,#2\right\rangle}
\def\spb#1.#2{\left[#1\,#2\right]}
\def\spash#1.#2{\spa{\smash{#1}}.{\smash{#2}}}
\def\spbsh#1.#2{\spb{\smash{#1}}.{\smash{#2}}}
\def\sand#1.#2.#3{%
\left\langle\smash{#1}{\vphantom1}^{-}\right|{#2}%
\left|\smash{#3}{\vphantom1}^{-}\right\rangle}
\def\sandpp#1.#2.#3{%
\left\langle\smash{#1}{\vphantom1}^{+}\right|{#2}%
\left|\smash{#3}{\vphantom1}^{+}\right\rangle}
\def\sandpm#1.#2.#3{%
\left\langle\smash{#1}{\vphantom1}^{+}\right|{#2}%
\left|\smash{#3}{\vphantom1}^{-}\right\rangle}
\def\sandmp#1.#2.#3{%
\left\langle\smash{#1}{\vphantom1}^{-}\right|{#2}%
\left|\smash{#3}{\vphantom1}^{+}\right\rangle}

\def\trFive{{\rm tr}_5}

\def\trFive{{\rm tr}_5}

\DeclareMathOperator{\codim}{codim}

\DeclareUnicodeCharacter{27E8}{\langle}
\DeclareUnicodeCharacter{27E9}{\rangle}
\DeclareUnicodeCharacter{0394}{\Delta}
\DeclareUnicodeCharacter{27EA}{\langle\langle}
\DeclareUnicodeCharacter{27EB}{\rangle\rangle}

\newcolumntype{C}[1]{>{\centering\arraybackslash}m{#1}}
\newcolumntype{M}{>{\centering\arraybackslash$}l<{$}}

\title{Primary Decompositions in Lorentz-Covariant Rings}

\preprint{}

\author[1]{Giuseppe~De~Laurentis,}
\author[1]{David Tai}

\affiliation[1]{Higgs Centre for Theoretical Physics, University of Edinburgh, Edinburgh, EH9 3FD, United Kingdom}

\emailAdd{
  giuseppe.delaurentis@ed.ac.uk,
  W.Tai-4@sms.ed.ac.uk,
  tai.wenzhao@gmail.com
}

\abstract{
    An effective strategy for constructing compact representations of integral coefficients in scattering amplitudes is to make their singularity structure manifest. This is naturally achieved by treating them as elements of fraction fields of polynomial quotient rings, where poles correspond to algebraic vanishing loci. Simplification via partial fraction decomposition relies on exploiting the structure of the numerators and is governed, at least in a first approximation, by their behaviour on irreducible codimension-two varieties. These are identifiable through primary decompositions of the associated ideals. Despite the availability of general algorithms, primary decompositions remain computationally challenging and often require tailored strategies.
    In this work, we extend an approach based on ideal saturation, identify an issue related to the breaking of Lorentz covariance in the choice of ideal generators, and present a solution based on fitting ansätze in special kinematic limits. We also provide substantial improvements to an algorithm for the generation of numerical points in proximity of given varieties and to one for testing primality of unmixed ideals, both of which rely on a semi-numerical procedure for computing independent sets and variety dimensions.
    Finally, we present new codimension-two primary decompositions for five-point massless amplitudes at higher powers in the dimensional regulator, and for finite remainders with five-point one-mass kinematics, where a new spurious singular locus appears.
}

\begin{document}
\maketitle

\section{Introduction}

State-of-the-art computations of scattering amplitudes relevant for
collider phenomenology involve two complementary computational
challenges: evaluating Feynman integrals with intricate analytic
structure, and managing rational coefficients with large algebraic
complexity.
In both cases, the structure of singularities plays a central
conceptual and computational role.
For Feynman integrals, these involve poles, branch points, and branch
cuts; see e.g.~\cite{Abreu:2022mfk} for a recent review.
For rational coefficients, singularities are encoded in the algebraic
pole loci of their denominators.

The study of singularities of Feynman integrals has a long history,
classically organised under the name of Landau analysis%
~\cite{Landau:1959fi,Nakanishi:1959jzx,Cutkosky:1960sp}.
Recent years have seen a resurgence of interest in this topic, driven
in part by reformulations in the language of computational algebraic
geometry%
~\cite{Panzer:2014caa,Mizera:2021icv,Klausen:2021yrt,
Fevola:2023kaw,Fevola:2023fzn},
by refinements based on stratifications%
~\cite{Helmer:2024wax,Helmer:2025ljj},
by connections to Grassmannian geometry and cluster structures%
~\cite{Hollering:2026gjz,Hollering:2026rlv},
and by public implementations such as \textsc{Sofia}%
~\cite{Caron-Huot:2024brh}, as well as related tools for polynomial
division and elimination such as
\textsc{SP}$\mathbb{Q}$\textsc{R}~\cite{Chestnov:2025svg}.
Beyond the determination of singular loci, Landau equations have also
found applications in constraining the analytic structure of Feynman
integrals \cite{Dennen:2015bet, Hannesdottir:2021kpd, Dlapa:2023cvx}
and in identifying the momentum regions relevant to asymptotic
expansions by the method of regions \cite{Ananthanarayan:2018tog,
  Gardi:2022khw, Gardi:2024axt}. The Landau equations form a system of
equations in the external and internal kinematic variables, where the
former describe the physical momenta of the scattered particles and
the latter the unobserved loop-integration degrees of freedom.
Eliminating the internal variables identifies Landau singular loci in
the space of external kinematics.  These are generally codimension-one
hypersurfaces and can have highly intricate algebraic
structure~\cite{Chestnov:2026mpo}.

It has been conjectured~\cite{Abreu:2018zmy} that these algebraic
singular loci are closely related to the possible denominator factors
appearing in the rational coefficients of scattering amplitudes.  This
conjecture can be understood in terms of the expected cancellations at
the vanishing loci of these denominators.  Since most denominator
factors are spurious---that is, they cancel in the full amplitude and
do not correspond to physical singularities on the first Riemann
sheet---the space spanned by the master integrals has to become
linearly dependent on these loci, thereby allowing such cancellations.
This is naturally explained if the vanishing loci of spurious
denominators are themselves special loci for the Feynman integrals.

The focus of this work is the geometric understanding of the singular
loci of rational coefficients.
Although the relevant computational methods are closely related to
those used for Feynman integrals, the two settings differ in an
important respect.
Most phenomenologically relevant amplitudes are gauge-theory
amplitudes and therefore involve polarisation degrees of freedom in
addition to the external momenta.
By contrast, the master integrals appearing in their decomposition are
Lorentz scalars.
Consequently, singularities of the integrals are naturally described
in terms of little-group-invariant Mandelstam variables, whereas their
coefficients generally carry little-group weights and are most
naturally expressed in spinor-helicity variables.

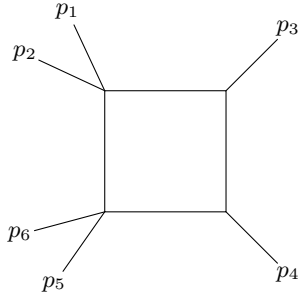
\begin{figure}[t]
    \centering
    \begin{tikzpicture}[scale=0.8, every node/.style={font=\small}, line cap=round]
        \draw (-1,1) -- (1,1) -- (1,-1) -- (-1,-1) -- cycle;

        \draw (-1,1)  -- ++(155:1.2);  
        \draw (-1,1)  -- ++(115:1.2);  

        \draw (1,1)   -- ++(45:1.2);   
        \draw (1,-1)  -- ++(-45:1.2);  

        \draw (-1,-1) -- ++(195:1.2);  
        \draw (-1,-1) -- ++(235:1.2);  

        \node at ($(-1,1)+(155:1.45)$)  {$p_2$};
        \node at ($(-1,1)+(115:1.45)$)  {$p_1$};

        \node at ($(1,1)+(45:1.45)$)    {$p_3$};
        \node at ($(1,-1)+(-45:1.45)$)  {$p_4$};

        \node at ($(-1,-1)+(195:1.45)$) {$p_6$};
        \node at ($(-1,-1)+(235:1.45)$) {$p_5$};
    \end{tikzpicture}
    \caption{Two-adjacent-mass box topology in six-point kinematics.}
    \label{fig:2am-box}
\end{figure}

This distinction has important algebraic consequences.  Polynomials
that are irreducible in a ring of Mandelstam variables may factor in
the corresponding spinor ring.  The simplest example is
$s_{12}=\langle12\rangle[21]$, but the phenomenon is more general.
For instance, another example is provided by the two-adjacent-mass box
integral in six-point kinematics shown in \cref{fig:2am-box}. Its Gram
determinant contains the non-trivial factor
\begin{equation*}
s_{12}(s_{123} - s_{56}) - s_{123}(s_{123} + s_{34} - s_{56}) = \langle 3\,|\,1{+}2\,|\,4]\langle4\,|\,1{+}2\,|\,3] \, .
\end{equation*}
It is often the case that only one of the two factors actually appears
in the denominators, because cancellations in the numerator can reduce
the order of the apparent pole.  A related phenomenon is that there
are quantities that are polynomials in spinors but square roots in
Mandelstams, such as
\begin{equation*}
\text{tr}_5^2 = \text{tr}(\gamma_5 p_1 p_2 p_3 p_4)^2 = ([1|2|3|4|1\rangle - \langle 1|2|3|4|1])^2 = \Delta_5 \, ,
\end{equation*}
where, in six-point kinematics,
\begin{equation*}
\begin{aligned}
    \Delta_5 = &\left(
        s_{123}(s_{234}-s_{34})
        +s_{23}(s_{34}-s_{56})
        +s_{12}(s_{23}-s_{234})
    \right)^2\\
    &\qquad+4s_{12}s_{23}s_{34}
    \left(
        s_{123}-s_{23}+s_{234}-s_{56}
    \right)
\end{aligned}
\end{equation*}
irreducible in the polynomial ring of Mandelstam variables.

More generally, the kinematic rings relevant for amplitude computations are not freely generated polynomial rings \cite{DeLaurentis:2022otd, Cederwall:2025ywy}. They are quotient rings, with relations imposed by momentum conservation and Schouten identities.\footnote{Relations among variables also play a role in integral computations, see for instance \cite{Abreu:2024fei}. The difference is that such relations set in at higher point for four-vectors compared to two-component spinors.} The resulting redundancies are often viewed as a computational
nuisance, and one may instead introduce a set of independent variables,
such as momentum twistors~\cite{Hodges:2009hk}.
While this makes some operations, such as differentiation,
significantly more straightforward, such changes of variables are
generally rational in nature, meaning that they turn polynomials into
rational functions. This obscures the structure of the singularities, generally increasing the degree of both numerators and denominators, and also hides factorization properties and symmetry relations.\footnote{For instance, at six points, one may write $s_{36}$ in terms of a minimal set of independent Mandelstam variables as $s_{12}-s_{123}-s_{345}+s_{45}$, which hides that it is in fact just a permutation of $s_{12}$.} For this reason, we are interested in studying these singularities in spinor space.

A further important motivation for this work is the study of multivariate
partial fraction decompositions (PFDs).
Their mathematical foundations go back to Leinartas'
decomposition%
~\cite{leinartas1978factorization,raichev2012leinartas};
see e.g.~\cite{Abreu:2019odu} for an early application to multiloop
amplitude coefficients.
Subsequent work developed improved algorithms and public
implementations, including an improved Leinartas algorithm for IBP
coefficients~\cite{Boehm:2020ijp},
\textsc{MultivariateApart}~\cite{Heller:2021qkz}, and
\textsc{pfd-parallel}~\cite{Bendle:2021ueg}.
More recently, criteria for the existence of desirable PFDs with poles
on hyperplane arrangements have been formulated in terms of primary
decompositions~\cite{deKorte:2026zrl}.
The kinematic setting considered here is more general, since the
denominator factors need not be linear and are elements of polynomial
quotient rings.
Nevertheless, the underlying geometric question is the same.
The codimension-one singular loci discussed so far, together with
their pole multiplicities, determine the least common denominator (LCD) of the rational coefficients, but expressions
written in LCD form are generally intractably complicated.
Their simplification depends on the geometry of the intersections among
these loci, which constrains the possible multivariate PFDs and, more
generally, the structure of the numerators%
~\cite{DeLaurentis:2022otd}.
Primary decompositions provide the algebraic tool for resolving these
intersections into irreducible branches.

Despite the availability of general algorithms, computing primary
decompositions in the Lorentz-covariant quotient rings relevant for
scattering amplitudes remains challenging.
For the high-dimensional ideals considered here, standard algorithms
often fail to terminate within a reasonable time, while intermediate
computer-algebra output need not be expressed in generators with
definite Lorentz transformation properties.
Further complications arise from non-radical primary components,
bottlenecks in generating points on varieties, and the difficulty of
testing whether ideals are prime or primary.

In this work, we develop a more systematic approach to these problems.
We reconstruct manifestly Lorentz-covariant generators by fitting
tensor ans\"atze on suitable algebraic varieties, and introduce
complementary analytic and numerical strategies for recovering
non-radical primary components from their associated primes.
We also improve the primality test for unmixed ideals and
extend it to primary ideals.
These developments rely in part on a semi-numerical procedure for
determining independent sets and variety dimensions, which also
substantially improves the generation of numerical points on
high-dimensional varieties.

We apply these methods to obtain new codimension-two primary
decompositions for five-point massless amplitudes involving
$\text{tr}_5$, and for five-point one-mass kinematics with denominator
structures appearing at two loops.
The former provide examples of non-radical primary components occurring
in decompositions with several associated primes, while the latter
extend the one-loop analysis of ref.~\cite[Appendix B]{DeLaurentis:2025dxw}.

The remainder of this paper is organised as follows.
In \cref{sec:fundamentals}, we review the relevant concepts from
algebraic geometry and introduce the Lorentz-covariant kinematic rings.
In \cref{sec:primary_and_partial_fraction_decompositions}, we explain the relation between primary
decompositions and multivariate partial fractions.
In \cref{sec:computing-primary-decompositions}, we describe our
computational approach, including numerical sampling, the reconstruction
of covariant generators, and the recovery of primary components from
their associated primes.
In \cref{sec:primalitytest}, we present the improved test for prime and
primary ideals.
Our new decompositions are given in \cref{sec:new-prim-dec}, with further details in \cref{5pt1mappendix}, and we
conclude in \cref{sec:conclusions}.

\section{Fundamental concepts and notation}
\label{sec:fundamentals}

In the hope of making this work accessible to both mathematicians
and theoretical physicists,
we begin by summarising the fundamental concepts
used throughout the rest of this work.
This section is divided into two parts.
Section~\ref{sec:alggeorev} reviews the algebro-geometric notions
relevant for primary decompositions of ideals.
Section~\ref{sec:kinemrev} recalls the kinematic variables and
algebraic relations that give rise to the polynomial quotient rings
encountered in scattering amplitudes.
The following section brings these two aspects together and explains
the relation between primary decompositions in Lorentz-covariant and
Lorentz-invariant quotient rings and partial fraction decompositions
of scattering amplitudes.

\subsection{Algebraic geometry definitions}\label{sec:alggeorev}
In this section, we briefly recall the notions of prime and primary ideals
and their relation to algebraic varieties through primary decompositions,
specialising to polynomial quotient rings relevant for kinematic applications.
This review sets the stage for the developments presented in the following sections.
A polynomial ring $S$ over a field $\mathbb{F}$ in $n$ variables
$\underline X=\{x_1,\dots,x_n\}$ is
\begin{equation}
    S=\mathbb{F}[\underline X]
    =
    \left\{
        \sum_{\alpha \in \mathbb{N}^n_0}
        c_\alpha x_1^{\alpha_1}\cdots x_n^{\alpha_n}
        \;\Big|\;
        c_\alpha\in\mathbb{F}
    \right\},
\end{equation}
consisting of all polynomials in the variables $\underline X$ with
coefficients in $\mathbb{F}$.
In other words, its elements are finite sums of monomials constructed
from these variables.
Together with the usual addition $(+)$ and multiplication $(\cdot)$ of
polynomials, $S$ is a ring, meaning that:
\begin{itemize}
    \item $(S,+)$ forms an abelian group with identity $0$;
    \item multiplication is associative and has identity $1$;
    \item multiplication distributes over addition:
    \(
        a(b+c)=ab+ac,\,
        (a+b)c=ac+bc
    \)
    for all $a,b,c\in S$.
\end{itemize}
A ring is \emph{commutative} if $ab=ba$ for all $a,b\in S$, and
\emph{Noetherian} if every ideal is finitely generated.
Polynomial rings over fields are both commutative and Noetherian.

An ideal $I\subset S$ is a non-empty subset that is closed under
addition and under multiplication by arbitrary elements of the ring:
\begin{itemize}
    \item if $f,g\in I$, then $f+g\in I$;
    \item if $f\in I$ and $r\in S$, then $rf\in I$.
\end{itemize}
Since $S$ is Noetherian, every ideal admits a finite set of generators
$f_1,\ldots,f_k$, and may be written as
\begin{equation}
    I=\langle f_1,\ldots,f_k\rangle
    =
    \left\{
        \sum_{i=1}^k r_i f_i
        \;\Big|\;
        r_i\in S
    \right\}.
\end{equation}
Thus, $I$ consists of all $S$-linear combinations of its generators,
that is, linear combinations whose coefficients are themselves
polynomials.
For example, the ideal $\langle x^2,xy\rangle$ contains
$x^2$, $xy$, $x^3=x\cdot x^2$, and
$x^2y+xy=y\cdot x^2+1\cdot xy$, together with infinitely many other
polynomials.

Several standard operations on ideals are useful in computational
applications:
\begin{itemize}
    \item \emph{Ideal sum}:
    \[
        I+J
        =
        \{f+g\mid f\in I,\ g\in J\}.
    \]
    If $I=\langle f_1,\ldots,f_m\rangle$ and
    $J=\langle g_1,\ldots,g_n\rangle$, then
    \[
        I+J
        =
        \langle f_1,\ldots,f_m,g_1,\ldots,g_n\rangle.
    \]

    \item \emph{Ideal product}:
    \[
        I \cdot J
        =
        \langle fg\mid f\in I,\ g\in J\rangle.
    \]

    \item \emph{Ideal intersection}:
    \[
        I\cap J
        =
        \{f\in S\mid f\in I\text{ and }f\in J\}.
    \]

    \item \emph{Ideal quotient}:
    \[
        I:J
        =
        \{f\in S\mid fg\in I\text{ for all }g\in J\}.
    \]

    \item \emph{Saturation}:
    \[
        I:J^\infty
        =
        \bigcup_{k\geq 1}(I:J^k)
        =
        \{f\in S\mid fJ^k\subseteq I
        \text{ for some }k\geq 1\}.
    \]
    Since $S$ is Noetherian, the ascending chain of ideal quotients
    eventually stabilises, and the smallest integer $s$ such that
    \[
        I:J^s=I:J^{s+1}
    \]
    is called the saturation index. One then has
    \(
        I:J^\infty=I:J^s.
    \)
\end{itemize}
Ideal quotients and saturations play an important role in primary
decomposition algorithms, where they can be used to isolate or remove
primary components, as in~\cite{DeLaurentis:2022otd}.

Given an ideal $J\subset\mathbb{F}[X]$, the quotient ring
\begin{equation}
    R=\mathbb{F}[X]/J
\end{equation}
is the set of equivalence classes of polynomials modulo $J$.
We denote an equivalence class by any of its representatives.
Thus, for $f,g\in\mathbb{F}[X]$, regarded as representatives of
elements of $R$,
\begin{equation}
    f=g\text{ in }R
    \quad\Longleftrightarrow\quad
    f-g\in J.
\end{equation}
This construction allows one to impose algebraic relations among the
variables, such as momentum conservation and on-shell conditions in
kinematic applications.

The variety $V(I)$ associated with an ideal $I$ is the set of common
zeros of all polynomials in $I$,
\begin{equation}
    V(I)
    =
    \left\{
        X\in\mathbb{F}^n
        \,\middle|\,
        f(X)=0
        \;\forall f\in I
    \right\}.
\end{equation}
Geometrically, $V(I)$ represents the solution set of the system of
polynomial equations defined by the generators of $I$. For an ideal $I\subset R=\mathbb{F}[X]/J$, $V(I)$ is understood as
the variety in $\mathbb{F}^n$ defined by $J$ together with
the generators of $I$, i.e.~$V(I+J)$.
Operations on ideals correspond to operations on varieties, as
summarised in~\cite[Section~4.8]{cox1994ideals}.

A proper ideal $P\subsetneq R$ is \emph{prime} if
\begin{equation}
    fg\in P
    \quad\Longrightarrow\quad
    f\in P
    \;\;\text{or}\;\;
    g\in P.
\end{equation}
Over an algebraically closed field, prime ideals correspond to
irreducible affine varieties, where \emph{irreducible} means that the
variety cannot be written as the union of two proper subvarieties.
More generally, a proper ideal $Q\subsetneq R$ is \emph{primary} if
\begin{equation}
    fg\in Q,\qquad f\notin Q
    \quad\Longrightarrow\quad
    g^k\in Q
    \quad\text{for some }k\geq1.
\end{equation}
The radical of an arbitrary ideal $I\subset R$ is defined by
\begin{equation}
    \sqrt{I}
    =
    \left\{
        f\in R
        \;\middle|\;
        f^k\in I
        \text{ for some }k\geq1
    \right\}.
\end{equation}
If $Q$ is primary, then its radical $P=\sqrt{Q}$ is prime and is
called the associated prime of $Q$.
The map $I\mapsto V(I)$ discards the multiplicity structure encoded in
a primary ideal: since
\(
    V(Q)=V(\sqrt{Q})=V(P),
\)
the variety alone cannot distinguish $Q$ from its associated prime.

\begin{figure}[htbp]
    \centering
    \begin{subfigure}[t]{0.40\textwidth}
        \centering
        \includegraphics[width=\textwidth]{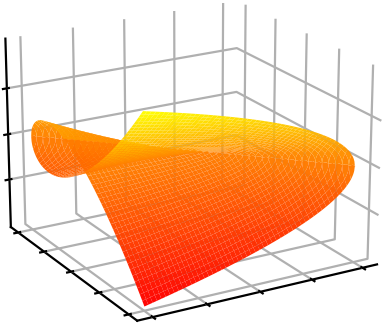}
        \caption{\centering An irreducible variety of dimension 2, $V(\langle xy^2 + y^3 - z^2\rangle)$.}
        \label{fig:variety_irreducible}
    \end{subfigure}
    \hspace{15mm}
    \begin{subfigure}[t]{0.40\textwidth}
        \centering
        \includegraphics[width=\textwidth]{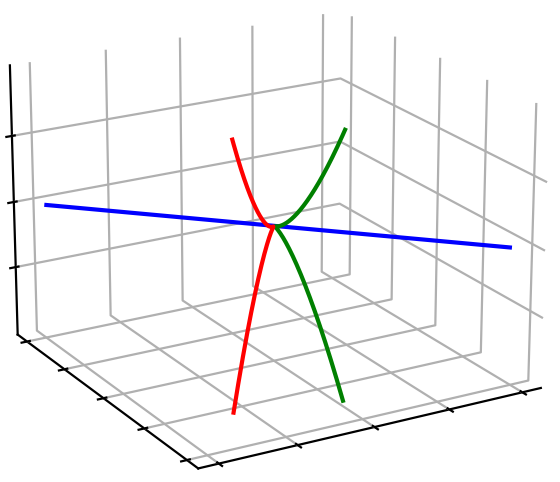}
        \caption{\centering A reducible variety of dimension 1, $V(\langle xy^2 + y^3 - z^2, x^3 + y^3 - z^2 \rangle)$.}
        \label{fig:variety_reducible}
    \end{subfigure}
    \caption{
    Geometric illustration of irreducible and reducible varieties.
    The variety on the left cannot be written as the union of two proper
    subvarieties, whereas the variety on the right is the union of the
    three irreducible varieties
    $V(\langle 2y^3-z^2,x-y\rangle)$,
    $V(\langle y^3-z^2,x\rangle)$, and
    $V(\langle z^2,x+y\rangle)$.
    Figures reproduced from Ref.~\cite{DeLaurentis:2022knk}.
    }
    \label{fig:varieties}
\end{figure}

Every ideal $I$ in a Noetherian ring admits a primary decomposition
\begin{equation}
    \label{eq:primary_decomp}
    I
    =
    \bigcap_{i=1}^{n_Q(I)} Q_i
    =
    Q_1\cap Q_2\cap\cdots\cap Q_{n_Q(I)},
\end{equation}
where each $Q_i$ is primary, with associated prime
$P_i=\sqrt{Q_i}$.
When the decomposition is minimal, meaning that no $Q_i$ is redundant
and the associated primes $P_i$ are distinct, the set of associated
primes is uniquely determined by $I$.
We denote it by
\begin{equation}
    \label{eq:assoc_primes}
    \operatorname{Assoc}(I)=\{P_i\}.
\end{equation}
The irreducible components of $V(I)$ are determined by the minimal
associated primes:
\begin{equation}
    V(I)
    =
    \bigcup_{P_i\in\operatorname{Min}(I)} V(P_i),
\end{equation}
where $\operatorname{Min}(I)\subseteq\operatorname{Assoc}(I)$ denotes
the set of minimal primes over $I$.
Any remaining associated primes are called embedded and do not define
additional irreducible components of the variety.
Figures~\ref{fig:variety_irreducible}
and~\ref{fig:variety_reducible} provide an intuitive illustration of
irreducible and reducible varieties.

In general, the irreducible components need not all have the same
dimension.\footnote{For example, the variety
\(
    V\bigl(\langle x(x-2),x(y+1)\rangle\bigr)
\)
has two irreducible components: the line $x=0$, of dimension one, and
the point $(x,y)=(2,-1)$, of dimension zero.}
If all irreducible components have the same dimension, then $I$ and
$V(I)$ are said to be \emph{equidimensional}.
This condition concerns only the minimal associated primes and does not
exclude embedded associated primes.
An ideal is instead called \emph{unmixed} if all its associated primes
have the same codimension.
A useful sufficient condition for unmixedness applies when the ambient
ring $R$ is Cohen--Macaulay.
If an ideal
\(
    I=\langle f_1,\ldots,f_c\rangle\subset R
\)
has maximal codimension,
\(
    \operatorname{codim}_R(I)=c,
\)
then $f_1,\ldots,f_c$ form a regular sequence, $R/I$ is
Cohen--Macaulay, and $I$ is unmixed
\cite[Proposition~18.13 and Corollary~18.14]{eisenbud1995commutative}.
In particular, $I$ has no embedded associated primes.
All primary decompositions computed in this work concern maximal-codimension ideals in Cohen--Macaulay ambient rings.

In Section~\ref{sec:primalitytest}, we make use of two additional
operations on ideals: \emph{extension} and \emph{contraction}.
Given the polynomial ring
\(
    \mathbb{F}[X]
\),
with
\(
    X=\{x_1,\ldots,x_n\},
\)
we split the variables into two disjoint sets $Y$ and $Z$, so that
$X=Y\sqcup Z$.
The \emph{extension} of an ideal
\(
    I=\langle p_1,\ldots,p_m\rangle\subset\mathbb{F}[X]
\)
is the ideal
\(
    I^e\subset\mathbb{F}(Y)[Z]
\)
defined by
\begin{equation}
    \label{eq:extension}
    I^e
    =
    I\,\mathbb{F}(Y)[Z]
    =
    \langle p_1,\ldots,p_m\rangle_{\mathbb{F}(Y)[Z]},
\end{equation}
where the subscript indicates the ambient polynomial ring.
Thus, $I^e$ is generated by the same generators as $I$, but these are
now viewed as polynomials in $Z$ whose coefficients are rational
functions in $Y$ rather than polynomials.

Conversely, the \emph{contraction} of an ideal
$J\subset\mathbb{F}(Y)[Z]$ is
\begin{equation}
    \label{eq:contraction}
    J^c=J\cap\mathbb{F}[X],
\end{equation}
that is, the subset of elements of $J$ that also belong to
$\mathbb{F}[X]$.
The composition of extension followed by contraction defines a map
$I\mapsto I^{ec}$ on the ideals of $\mathbb{F}[X]$, with
$I\subseteq I^{ec}$ in general.
It can be shown that $I^{ec}$ can be computed as the saturation
\begin{equation}
    \label{eq:extension-contraction-through-saturation}
    I^{ec}=I:f^\infty,
\end{equation}
where $f$ is the least common multiple of the head coefficients of a
Gr\"obner basis $\mathcal{G}$ of $I$, computed in a block ordering with
$Z\succ Y$~\cite{becker2012groebner,DeLaurentis:2022otd}.
Here each $g\in\mathcal{G}$ is viewed as a polynomial in $Z$ with
coefficients in $\mathbb{F}[Y]$, and
\begin{equation}
    \label{eq:fpoly}
    f
    =
    \operatorname{lcm}
    \left\{
        \operatorname{HC}_Z(g)\in\mathbb{F}[Y]
        \,\middle|\,
        g\in\mathcal{G}
    \right\}.
\end{equation}
Geometrically, extension followed by contraction removes the primary
components $Q_i$ for which $f\in\sqrt{Q_i}$, equivalently those whose
associated varieties are contained in $V(f)$.
It plays a central role in the primality test described in
Section~\ref{sec:primalitytest}.

\subsection{Lorentz-covariant and Lorentz-invariant rings}\label{sec:kinemrev}

The spinor-helicity formalism provides a natural framework for
describing massless scattering amplitudes.
Each massless four-momentum $p_i$ is encoded by a pair of Weyl spinors,
$\lambda_i^\alpha$ and $\tilde\lambda_i^{\dot\alpha}$,
\begin{equation}
    p_i^{\alpha\dot\alpha}
    =
    \lambda_i^\alpha\tilde\lambda_i^{\dot\alpha}.
\end{equation}
This rank-one representation automatically satisfies the on-shell
condition $p_i^2=0$.

Following Ref.~\cite{DeLaurentis:2022otd}, we describe the kinematic
space using polynomial quotient rings constructed from the spinor
components.
The component polynomial ring is
\begin{equation}\label{eq:component_polynomial_ring}
 S_n = \mathbb{F}\big[ \lambda^\alpha_1, \tilde\lambda^{\dot\alpha}_1, \dots, \lambda^\alpha_n, \tilde\lambda^{\dot\alpha}_n \big]\,,
\end{equation}
which contains all polynomials in the $4n$ scalar components of the
spinors.
Momentum conservation imposes the four polynomial relations
\begin{equation}
    \sum_{i=1}^n
    \lambda_i^\alpha\tilde\lambda_i^{\dot\alpha} = 0,
\end{equation}
which generate the momentum-conservation ideal
\begin{equation}
    J_n =
    \left\langle
        \sum_{i=1}^n
        \lambda_i^\alpha\tilde\lambda_i^{\dot\alpha}
        \,\middle|\,
        \alpha,\dot\alpha=1,2
    \right\rangle
    \subset S_n.
\end{equation}
The corresponding momentum-conserving component quotient ring is therefore
\begin{equation}
    \label{eq:lorentz_covariant_ring}
    R_n = S_n/J_n.
\end{equation}

We refer to $R_n$ as the \emph{Lorentz-covariant ring}.
The ring contains arbitrary polynomials in the spinor components, which
need not themselves have definite Lorentz transformation properties.
In the applications below, however, we will be interested principally
in elements that transform covariantly in definite Lorentz
representations.

Lorentz-invariant\footnote{By ``Lorentz invariant'' we mean invariant
under the independent actions of $SL(2)_L$ and $SL(2)_R$ on undotted
and dotted spinor indices, respectively. Little-group scalings are
kept separate.} polynomials are generated by contracting the spinors
into angle and square brackets. In our chosen convention, these are defined by
\begin{equation}
    \langle ij\rangle = \lambda_i^\alpha\lambda_{j\alpha},
    \qquad
    [ij] = \tilde\lambda_{i\dot\alpha} \tilde\lambda_j^{\dot\alpha},
\end{equation}
where spinor indices are raised and lowered using the appropriate 
anti-symmetric tensor, $\epsilon^{\alpha\beta}=\epsilon^{\dot\alpha\dot\beta}=\epsilon$ and  $\epsilon_{\alpha\beta}=\epsilon_{\dot\alpha\dot\beta}=\epsilon^T$.
In parallel with the component ring $S_n$, the bracket polynomial ring
is defined as
\begin{equation}
    \mathcal{S}_n
    =
    \mathbb{F}
    \big[
        \langle ij\rangle,[ij]
        \,\big|\,
        1\leq i<j\leq n
    \big].
\end{equation}
At this stage, the brackets are treated as formal polynomial variables,
with brackets in the opposite ordering defined by antisymmetry.
Their algebraic relations are imposed through a quotient.

Contracting momentum conservation with
$\langle j|$ and $|k]$ gives
\begin{equation}
    \sum_{i=1}^n
    \langle ji\rangle[ik]
    =
    0,
    \qquad
    j,k=1,\ldots,n,
\end{equation}
and hence the momentum-conservation ideal
\begin{equation}
    \mathcal{J}^{\mathrm{mom}}_n
    =
    \left\langle
        \sum_{i=1}^n\langle ji\rangle[ik]
        \,\middle|\,
        j,k=1,\ldots,n
    \right\rangle_{\mathcal{S}_n}.
\end{equation}
The angle and square brackets also satisfy the Schouten relations,
which generate the ideals
\begin{align}
    \mathcal{K}^{\langle\,\rangle}_n
    &=
    \big\langle
        \langle ij\rangle\langle k\ell\rangle
        -
        \langle ik\rangle\langle j\ell\rangle
        +
        \langle i\ell\rangle\langle jk\rangle
        \,\big|\,
        1\leq i<j<k<\ell\leq n
    \big\rangle_{\mathcal{S}_n},
    \\
    \mathcal{K}^{[\,]}_n
    &=
    \big\langle
        [ij][k\ell]
        -
        [ik][j\ell]
        +
        [i\ell][jk]
        \,\big|\,
        1\leq i<j<k<\ell\leq n
    \big\rangle_{\mathcal{S}_n}.
\end{align}
These identities follow from the fact that any three vectors in a
two-dimensional space are linearly dependent.
Together with the momentum-conservation relations, the angle- and
square-bracket Schouten relations generate the complete ideal of
kinematic relations,
\begin{equation}
    \mathcal{J}_n
    =
    \mathcal{J}^{\mathrm{mom}}_n
    +
    \mathcal{K}^{\langle\,\rangle}_n
    +
    \mathcal{K}^{[\,]}_n,
\end{equation}
and the corresponding Lorentz-invariant quotient ring is
\begin{equation}
    \label{eq:lorentz_invariant_ring}
    \mathcal{R}_n
    =
    \mathcal{S}_n/\mathcal{J}_n.
\end{equation}
The natural map replacing the formal bracket variables by their
spinor-component expressions induces an isomorphism between
$\mathcal{R}_n$ and the Lorentz-invariant subring of $R_n$, as shown in
Ref.~\cite{DeLaurentis:2022otd}.

Although scattering amplitudes are Lorentz invariant, their singular
loci can often be described more concisely in the
Lorentz-covariant ring $R_n$ than in the invariant ring
$\mathcal{R}_n$.
Open spinor indices allow several scalar generators to be packaged into
a single covariant tensor generator.
For example, at $n>4$ points, consider the locus on which the angle
spinor $|i\rangle$ vanishes.
In $R_n$, it is described by the covariant ideal
\begin{equation}
    I
    =
    \big\langle |i\rangle\big\rangle_{R_n},
\end{equation}
where an ideal generated by a covariant tensor is understood to be
generated by all its spinor components.
Its contraction to the Lorentz-invariant subring is
\begin{equation}
    I\cap\mathcal{R}_n
    =
    \big\langle
        \langle ji\rangle
        \,\big|\,
        j\neq i
    \big\rangle_{\mathcal{R}_n}.
\end{equation}
Thus, the same locus is encoded by a single covariant spinor generator
in $R_n$, but by the $n-1$ non-trivial invariant contractions involving
that spinor in $\mathcal{R}_n$.
This motivates performing primary decompositions in the
Lorentz-covariant ring and expressing their components in terms of
generators with definite Lorentz transformation properties.

A Lorentz-invariant polynomial $f(\lambda,\tilde\lambda)$ satisfies
\begin{equation}
    \label{eq:Lorentz_scalar_function}
    f(\Lambda\lambda,\tilde\Lambda\tilde\lambda)
    =
    f(\lambda,\tilde\lambda),
\end{equation}
where $\Lambda\in SL(2)_L$ and
$\tilde\Lambda\in SL(2)_R$ act on the undotted and dotted spinors,
respectively.
Here Lorentz invariance does not imply little-group invariance: under
particle-wise little-group scalings, such a polynomial may carry
non-zero homogeneous weights.
By contrast, a Lorentz-covariant polynomial carrying $r$ undotted and
$s$ dotted indices transforms as
\begin{align}
    &f^{\alpha_1\cdots\alpha_r
       \dot\alpha_1\cdots\dot\alpha_s}
    (\Lambda\lambda,\tilde\Lambda\tilde\lambda)
    \nonumber\\
    &\qquad =
    \Lambda^{\alpha_1}{}_{\beta_1}\cdots
    \Lambda^{\alpha_r}{}_{\beta_r}
    \tilde\Lambda^{\dot\alpha_1}{}_{\dot\beta_1}\cdots
    \tilde\Lambda^{\dot\alpha_s}{}_{\dot\beta_s}
    f^{\beta_1\cdots\beta_r
       \dot\beta_1\cdots\dot\beta_s}
    (\lambda,\tilde\lambda).
    \label{eq:covariant_generator}
\end{align}

In practical computations, covariant generators are represented by
their individual spinor components, reducing the problem to scalar
polynomial algebra.
Symbolic algebra systems such as \textsc{Singular}, however, do not
retain the Lorentz transformation properties that organise these
components into tensors.
Consequently, standard primary-decomposition algorithms and ideal
operations such as saturation may return generators that do not have
definite Lorentz transformation properties.
For example, an output polynomial may take the form $f=g+h$, where
$g$ and $h$ belong to different Lorentz representations, such as an
invariant polynomial and one component of a covariant tensor.
Additional techniques are therefore required to separate these
contributions and reconstruct generators with definite Lorentz
transformation properties.

\section{Primary and partial-fraction decompositions}
\label{sec:primary_and_partial_fraction_decompositions}

Scattering amplitudes in quantum field theory encode the probability
for a given set of final-state particles to emerge from the interaction
of particles in the initial state.
Working in dimensional regularisation, after ultraviolet renormalisation
and subtraction of infrared singularities, an $n$-point
$\ell$-loop amplitude, $\mathcal{A}^{(\ell)}_n$, can be represented by
a finite remainder, $\mathcal{F}^{(\ell)}_n$.
The latter can be expressed as a linear combination of transcendental
functions $H_i$\footnote{For convenience, we include the constant
function $1$ among the $H_i$.}, with rational-function coefficients
$C_i$,
\begin{equation}\label{eq:amplitude_decomposition}
    \mathcal{F}^{(\ell)}_n
    =
    \sum_i
    C_i(\lambda,\tilde\lambda)\,
    H_i(\lambda\tilde\lambda).
\end{equation}
In gauge theories, the coefficients $C_i$ depend separately on the
holomorphic and anti-holomorphic spinor-helicity variables, while the
functions $H_i$ depend only on the momenta
$p_i=\lambda_i\tilde\lambda_i$.
We focus on the rational coefficients.
They can be written in least-common-denominator (LCD) form as
\begin{equation}\label{eq:rational_coefficient}
    C_i(\lambda,\tilde\lambda)
    =
    \frac{N_i(\lambda,\tilde\lambda)}
    {\prod_j D_j(\lambda,\tilde\lambda)^{q_j}},
\end{equation}
where the $N_i$ and $D_j$ are polynomials in spinor brackets with
well-defined mass dimensions and little-group weights, and
$q_j\in\mathbb Z$ specifies the order of the pole for $q_j>0$ or of
the zero for $q_j<0$ associated with $D_j$.
The LCD form is understood to be reduced, so that no factor with
$q_j>0$ can be cancelled against the numerator $N_i$.
A central challenge in modern amplitude calculations is to determine
these rational coefficients, and an effective approach is to understand
the locations and orders of their singularities.

Each denominator factor $D_j$ defines a codimension-one variety
$V(\langle D_j\rangle)$ in the corresponding kinematic space associated
with $R_n$ (or $\mathcal{R}_n$).
These loci may correspond to physical singularities, such as those
arising in soft and collinear limits, or to spurious singularities,
which cancel in the finite remainder $\mathcal{F}^{(\ell)}_n$.
The behaviour of the $C_i$ on these codimension-one varieties determines
the exponents $q_j$ and hence their LCD form.
However, the numerator $N_i$ in LCD form is often intractably
complicated.
It is therefore useful to consider partial-fraction representations of
the rational coefficients.

A partial-fraction decomposition seeks to rewrite
\cref{eq:rational_coefficient} as a sum of simpler terms, typically
with lower-degree numerators and denominators,
\begin{equation}\label{eq:partial_fraction_decomposition}
    C_i(\lambda,\tilde\lambda)
    =
    \sum_k
    \frac{N_{ik}(\lambda,\tilde\lambda)}
    {\prod_j D_j(\lambda,\tilde\lambda)^{q_{jk}}}.
\end{equation}
Such a representation is free of spurious poles if, for all $j$ and $k$,
$q_{jk}\leq q_j$ when $q_j>0$, and $q_{jk}\leq0$ when $q_j\leq0$,%
\footnote{Here, a spurious pole is understood relative to the individual
coefficient $C_i$, rather than to the complete finite remainder
$\mathcal{F}^{(\ell)}_n$.}
although partial-fraction decompositions need not in general satisfy
this condition.
Just as the LCD form reflects the behaviour of $C_i$ at codimension
one, a partial-fraction representation is sensitive to its behaviour
at higher codimension.
In practice, a substantial amount of this information is already
accessible at codimension two.

When two denominator factors $D_a$ and $D_b$ vanish simultaneously,
their intersection
\begin{equation}
    V(\langle D_a,D_b\rangle)
    =
    V(\langle D_a\rangle)\cap V(\langle D_b\rangle)
\end{equation}
is generically of codimension two.
The ideal $\langle D_a,D_b\rangle$ need not be prime, and its primary
decomposition
\begin{equation}\label{eq:primdecDaDb}
    \langle D_a,D_b\rangle
    =
    Q_1\cap Q_2\cap\cdots\cap Q_m
\end{equation}
resolves the intersection into its irreducible branches, described by
the associated primes $P_k=\sqrt{Q_k}$.
The rational coefficients need only have uniform behaviour at generic
points on each branch, and not necessarily across different branches.

For simple poles in $D_a$ and $D_b$, a partial-fraction decomposition
that separates the two factors exists precisely when the LCD numerator
satisfies
\begin{equation}
    N_i\in\langle D_a,D_b\rangle
    \quad\Longleftrightarrow\quad
    N_i=A\,D_a+B\,D_b
\end{equation}
for some polynomials $A$ and $B$.
Using the primary decomposition of \cref{eq:primdecDaDb},
this condition is equivalently imposed on every primary component,
$N_i\in Q_k$ for all $k$.
When the ideal $\langle D_a,D_b\rangle$ is radical, Hilbert's
Nullstellensatz gives the equivalent geometric statement that
$N_i$ vanishes on every irreducible branch,
\begin{equation}
    N_i
    \in
    \sqrt{\langle D_a,D_b\rangle}
    =
    P_1\cap\cdots\cap P_m.
\end{equation}
Thus, the primary decomposition determines the individual
codimension-two branches on which the numerator constraints relevant
to partial fractions must be studied
\cite{DeLaurentis:2022otd,Campbell:2022qpq,deKorte:2026zrl}.

More generally, for higher-order poles, a partial-fraction decomposition
requires not only determining whether the numerator vanishes on each
irreducible branch, but also its order of vanishing.
The Zariski--Nagata theorem provides a higher-order analogue of
Hilbert's Nullstellensatz, relating vanishing to order $\kappa$ along
$V(P_k)$ to membership in the $\kappa$th symbolic power
$P_k^{\langle\kappa\rangle}$~\cite{DeLaurentis:2022otd}.
More generally, the symbolic power may be viewed as the component of
the ordinary power associated with the same irreducible variety,
excluding any additional embedded components that may arise.

When an ideal is generated by a regular sequence, its symbolic and
ordinary powers coincide.
In particular, the maximal-codimension ideal
$\langle D_a,D_b\rangle$ is generated by a regular sequence in the
Cohen--Macaulay rings considered here~\cite{DeLaurentis:2022otd}, and
therefore
\begin{equation}
    \langle D_a,D_b\rangle^{\langle\kappa\rangle}
    =
    \langle D_a,D_b\rangle^\kappa .
\end{equation}
This simplification applies to the full intersection ideal; the
individual associated primes $P_k$ need not themselves be complete
intersections.
Thus, symbolic powers provide the natural extension of the
codimension-two analysis above when higher-order vanishing conditions
are relevant.

\subsection{Irreducible denominator factors}\label{sec:irred_denoms}

We now specify the irreducible denominator factors $D_a$ from which the
codimension-two ideals studied in this work are constructed.
In the following, denominator types are given up to permutations of
kinematically equivalent external legs and parity conjugation,
$\langle\,\rangle\leftrightarrow[\,]$.

For five-point massless kinematics, the denominator factors appearing in
finite remainders through two loops are generated by just two
representatives,
\begin{equation}\label{eq:five-point-massless-invariants}
    D_a
    \sim
    \big\{
        \langle12\rangle,\,
        \langle1|2{+}3|1]
    \big\},
\end{equation}
where $\sim$ denotes equivalence under permutations of the five external
legs and parity conjugation.
Spinor chains are defined, for example, by
\begin{equation}
    \langle i|j{+}k|l]
    =
    \langle ij\rangle[jl]
    +
    \langle ik\rangle[kl],
\end{equation}
so that, in particular,
$\langle i|j{+}k|i]=s_{ij}+s_{ik}$, with
$s_{ij}=\langle ij\rangle[ji]$.
The complete set of codimension-two primary decompositions constructed
from the denominator types in \cref{eq:five-point-massless-invariants}
was obtained in Ref.~\cite{DeLaurentis:2022otd}.
When the amplitudes are expanded to higher orders in the dimensional
regulator $\epsilon$, the parity-odd invariant
\begin{equation} \label{eq:tr_5-definition}
    \operatorname{tr}_5(i,j,k,\ell)
    =
    4i\,\varepsilon^{\mu\nu\rho\sigma}
    p_{i,\mu}p_{j,\nu}p_{k,\rho}p_{\ell,\sigma}
    =
    [i|j|k|\ell|i\rangle
    -
    \langle i|j|k|\ell|i]
\end{equation}
also appears as a denominator factor.
The relevant set of representatives is therefore enlarged to
\begin{equation}\label{eq:five-point-massless-invariants-higher-eps}
    D_a
    \sim
    \big\{
        \langle12\rangle,\,
        \langle1|2{+}3|1],\,
        \operatorname{tr}_5(1,2,3,4)
    \big\}.
\end{equation}
In \cref{sec:res-five-point-massless}, we complete the massless
five-point analysis by computing the two inequivalent codimension-two
decompositions involving $\operatorname{tr}_5$.
Both contain non-radical primary components, and together they introduce
one new associated prime.

For five-point one-mass kinematics, we label the four massless momenta
by $p_1,\ldots,p_4$ and the massive momentum by $p_{\boldsymbol{5}}$.
The irreducible denominator factors observed in the known two-loop
leading-colour finite remainders are, up to permutations of the four
massless legs and parity conjugation, represented by
\begin{equation}\label{eq:five-point-one-mass-invariants}
    D_a
    \sim
    \big\{
        \langle12\rangle,\,
        \langle1|2{+}3|1],\,
        \langle1|2{+}3|4],\,
        s_{123},\,
        \Delta_{12|34|\boldsymbol{5}},\,
        \langle3|2|\boldsymbol{5}|4|3]
        -
        \langle2|1|\boldsymbol{5}|4|2]
    \big\}.
\end{equation}
Here
\begin{equation}
    s_{ijk}
    =
    (p_i+p_j+p_k)^2
    =
    s_{ij}+s_{ik}+s_{jk}
\end{equation}
for massless $i,j,k$.
The Gram determinant is defined, for $K_1+K_2+K_3=0$, by
\begin{equation}
    \Delta_{K_1|K_2|K_3}
    =
    (K_1\cdot K_2)^2-K_1^2K_2^2
    =
    \frac{1}{4}
    \left[
        \bigl(s_{K_1K_2}-K_1^2-K_2^2\bigr)^2
        -
        4K_1^2K_2^2
    \right],
\end{equation}
where $s_{K_1K_2}=(K_1+K_2)^2$.
Thus, $\Delta_{12|34|\boldsymbol{5}}$ corresponds to
$K_1=p_1+p_2$, $K_2=p_3+p_4$, and
$K_3=p_{\boldsymbol{5}}$.

Following Ref.~\cite{DeLaurentis:2025dxw}, in practical computations
the massive momentum can be represented as the sum of two massless momenta,
$p_{\boldsymbol{5}}=p_5+p_6$, so that the kinematics are embedded in
$R_6$.
Spinor chains involving the massive leg can then be expanded in terms
of the auxiliary massless spinors; for example,
\begin{equation}
    \langle i|j|\boldsymbol{5}|k|i]
    =
    \langle i|j|5]\langle5|k|i]
    +
    \langle i|j|6]\langle6|k|i].
\end{equation}
Among the representative denominator types in
\cref{eq:five-point-one-mass-invariants}, only the final
mass-dimension-four difference of spinor chains is new at two loops.
Ref.~\cite{DeLaurentis:2025dxw} provided the primary decompositions for
all denominator-pair ideals constructed from the one-loop factors,
together with a selected decomposition involving the new two-loop
factor.
In \cref{sec:res-five-point-onemass}, we complete this analysis by
computing the inequivalent codimension-two primary decompositions
involving the new denominator type, resulting in 17 new associated
primes.

Interestingly, the same denominator types have also been observed when
non-planar contributions are present.
The leading-colour five-point one-mass amplitudes of
Ref.~\cite{DeLaurentis:2026brm} contain non-planar contributions, but
their rational coefficients exhibit no additional irreducible
denominator factors beyond those already present in the planar
five-point one-mass amplitudes.
Whether this remains true for two-loop five-point one-mass finite
remainders at full colour is not yet established.
In principle, this question could already be investigated using the
full-colour results of Ref.~\cite{Badger:2024mir}.
Doing so, however, requires several non-trivial steps, including
converting the rational functions from momentum-twistor to
spinor-helicity variables, assembling the helicity amplitudes from the
form factors, and recomputing their LCDs.

\section{Computing primary decompositions}
\label{sec:computing-primary-decompositions}

The main problem we aim to address is the computation of primary decompositions, as defined in \cref{eq:primary_decomp,eq:assoc_primes}. We focus on the case of ideals of high dimensions (the results presented in \cref{sec:res-five-point-massless,sec:res-five-point-onemass} have dimension $14$ and $18$, respectively) whose primary components have uniform dimension. Automated algorithms such as those by  Gianni, Trager and Zacharias \cite{gianni1988grobner}, and by Shimoyama and Yokoyama \cite{SHIMOYAMA1996247}, which are implemented in \textsc{Singular}, in principle offer a general solution to this problem. However, for the ideals under consideration,
these methods often fail to terminate within a reasonable time.

In this work, we improve upon the approach to primary decomposition presented in ref.~\cite{DeLaurentis:2022otd}, addressing issues that bottleneck the automation of this approach. In fact, decompositions such as those computed in refs.~\cite{DeLaurentis:2022otd} and \cite{DeLaurentis:2025dxw} were obtained partly by hand, while also not covering the full range of features and complexity accessible in similar quantum field theory computations. 

We describe here steps towards a semi-numerical algorithm that can effectively address this problem for Lorentz-covariant ideals of uniform dimension.

\subsection{Sampling solutions to polynomial systems}\label{sec:variety_sampling}

The first step towards simplifying the primary decomposition problem
is to identify known primary components, which can then be removed via saturation, a comparatively inexpensive operation in this setting. An efficient way to identify components against a set of known ones is through numerical sampling, which can be done reliably in a finite field. In ref.~\cite[Section 3.2]{DeLaurentis:2022otd} an algorithm to perform this sampling was presented. It can be summarized as follows. Start from an ideal
\begin{equation}
J = \big \langle q_1(\underline X) \dots q_m(\underline X) \big\rangle_{\mathbb{F}[\underline X]} \, ,
\end{equation}
and pick an independent set $\underline Y$, with corresponding dependent variables $\underline Z=\underline X \backslash \underline Y$, where
\begin{equation}
\text{dim}(\underline Y) = \text{dim}(J) \, \text{ and } \, \text{dim}(\underline Z) = \text{codim}(J) \, .
\end{equation}

Let $\underline Y^{(0)} \in \mathbb{F}^{\dim(\underline Y)}_p$
be a random set of values for $\underline Y$.
By specialising the variables $\underline Y$ to these values,
that is, by applying the substitution
$\underline Y \mapsto \underline Y^{(0)}$,
we obtain a zero-dimensional slice of $V(J)$,
described by the ideal\footnote{
If $J$ is an ideal of a quotient ring $\mathbb{F}[\underline X]/K$,
we instead work with the ideal $J+K$ in $\mathbb{F}[\underline X]$.
}
\begin{equation}
J^{(0)}
=
\big\langle
q_1(\underline Z,\underline Y^{(0)}),
\dots,
q_m(\underline Z,\underline Y^{(0)})
\big\rangle_{\mathbb{F}[\underline Z]} \, .
\end{equation}
The points comprising $V(J^{(0)})$ can be obtained
by successively solving univariate polynomial equations
in the lexicographic Gr\"obner basis of $J^{(0)}$.
We denote them by
$\underline Z^{(0)}_1,\dots,\underline Z^{(0)}_n$.\footnote{
If no solution exists in $\mathbb{F}_p$,
it suffices to repeat the procedure for a different random
$\underline Y^{(0)}$.
}
Points $\underline X^{(0)}_1,\dots,\underline X^{(0)}_n$
on $V(J)$ are then obtained by adjoining
$\underline Y^{(0)}$ to each $\underline Z^{(0)}_i$.

The first step of this procedure, namely the computation of an independent set $\underline Y$, suffers from dependency on the complexity of $J$, including its dimension and the complexity of the generators $q_i$. In practice, \textsc{Singular}'s \texttt{indepSet} function requires a standard basis (read Gr\"oebner basis) of $J$, which may not be computable in a reasonable time. The solution we propose and implement is to determine the independent set $\underline Y$ semi-numerically. As a byproduct, the same procedure also yields $\text{dim}(J)$, which is a bottleneck for the primality test discussed in the next section. 

First, note that the number of generators, $m$, provides an upper bound on the codimension 
\begin{equation}
    \text{codim}(J) \leq m \, ,
\end{equation}
or equivalently a lower bound for the dimension. If the bound is saturated, i.e.~if $\text{codim}(J) = m$, then $J$ is said to be of maximal codimension. We take a random subset $\underline W$ of $\underline X$ with 
\begin{equation}
    \text{dim}(\underline W) = \text{dim}(\underline X) - m \, .
\end{equation} 
Then, we construct the ideal 
\begin{equation}
J' = \big \langle q_1(\underline Z', \underline W^{(0)}) \dots q_m(\underline Z', \underline W^{(0)}) \big\rangle_{\mathbb{F}[\underline Z']} \, .
\end{equation}
for $\underline Z' = \underline X \backslash \underline W$ and a generic random sample $\underline W^{(0)} \in \mathbb{F}^{\dim(W)}_p$. One of two things can happen:
\begin{enumerate}
    \item[a.] If there exists an independent set $\underline Y \supseteq \underline W$, then
    \begin{equation}
        \text{dim}(J') = \text{dim}(\underline Y) - \text{dim}(\underline W) \geq 0 \, .
    \end{equation}
    \item[b.] If no such independent set exists, then $J'=\big\langle 1 \big\rangle$ and hence $V(J') = \emptyset$.
\end{enumerate}
To determine in which case we land, it suffices to compute the dimension of $J'$. This is usually much simpler than computing the dimension or an independent set of $J$, since the variables in $\underline W$ no longer appear as variables
of the polynomial ring, having been specialised to values in the field.

If we landed in case b., then we cannot determine the dimension of $J$ and we have to repeat the procedure until we hit case a. Once we hit case a, then we have\footnote{There exists a (small) chance for points thus constructed to land on a sub-variety of $V(J)$ which has lower dimension than $V(J)$, e.g. in the case where $J$ has embedded components. This can be recognized and accommodated for by generating a sufficiently large set of generic points on $V(J)$.}
\begin{equation}
\text{dim}(J)=\text{dim}(\underline Y)=\text{dim}(J') + \text{dim}(\underline W) \, .
\end{equation}
We can then proceed with the algorithm to determine points on $V(J)$ by picking random sets $\underline W$ of the correct dimension, i.e.~that either saturate the bound in case a, and hence are independent sets $\underline Y = \underline W$, or that are not independent sets and land in case b. This approach provides orders of magnitude speed ups for a variety of cases tested. 

Lastly, if the computation of $\dim(J')$ itself becomes too expensive,
for instance in cases where $m \gg \codim(J)$,
one may use a more aggressive guessing strategy.
Instead of using the general bound $\codim(J)\le m$,
we choose a smaller guessed upper bound $c<m$ for the codimension
and take
\begin{equation}
    \dim(\underline W)=\dim(\underline X)-c \, .
\end{equation}
If the guess still satisfies $c\ge \codim(J)$,
then $\underline W$ may be contained in an independent set,
and the procedure can land in case a.
If, however, $c<\codim(J)$,
then $\dim(\underline W)>\dim(J)$,
so no independent set of $J$ can contain $\underline W$.
Consequently, the procedure can never land in case~a,
regardless of how many random specialisations are attempted.
In this situation one can progressively relax the guess,
increasing $c$ until case~a is reached.\footnote{
We implement this in \textsc{syngular},
\texttt{ideal.point\_on\_variety},
and provide an attribute \texttt{ideal.codim\_upper\_bound}
which can be used to replace the default bound $m$
when guessing independent sets
\cite{syngular}.
}
 
With this improved procedure at hand,
we are able to reliably sample, within reasonable computational time,
hundreds to thousands of points on each of the varieties under consideration.
Crucially, the sampling is generic and random.
If the ideal $J$ is not primary and admits a (yet to be determined)
primary decomposition
\begin{equation}
J = \bigcap_{i=1}^{n_Q(J)} Q_i = Q_1 \cap \dots \cap Q_{n_Q(J)} \,,
\end{equation}
then the sampled points $\underline X^{(0)}$
will typically be distributed across the various components $V(Q_i)$.
Although this distribution is not uniform,
we observe empirically that,
provided sufficiently many points are generated,
all components are eventually sampled.

In many applications, additional structural information is available.
In particular, one may have a distinguished collection of polynomials
\be
r = \{r_1,\dots,r_k\} \subset R
\ee
whose vanishing defines geometrically meaningful loci.
In the scattering-amplitude setting,
these typically correspond to kinematic poles $D_a$ of the amplitude coefficients.
We exploit this additional structure to guide the identification of candidate associated primes.

For each sampled point $\underline X^{(0)}_{\,j} \in V(J)$,
we evaluate the distinguished polynomials
and record the subset that vanishes at the point
\begin{equation}
v_j = \{\, r_i \mid r_i(\underline X^{(0)}_j) = 0 \,\} \,.
\end{equation}
This produces a combinatorial signature
encoding which point landed on which variety. 
For sufficiently large sets $\underline r$, the number of distinct
signatures typically agrees with the number of associated primes,
or equivalently with the number of primary components in a minimal
decomposition:
\[
    \#\{v_j\} = n_Q(J).
\] In many cases, but not always, the polynomials comprising the $v_j$ are (non minimal) generating sets of the corresponding ideals.

Independently, consider a list of candidate prime ideals
$P_\alpha$ with irreducible varieties $V(P_\alpha)$.
Such candidates may arise, for instance,
from earlier computations involving subsets
of the distinguished polynomials.
For each such prime,
a single generic point $\underline X^{(0)}_\alpha$
suffices to determine its associated vanishing set
\begin{equation}
\mathcal{V}(P_\alpha)
=
\{\, r_i \mid r_i(\underline X^{(0)}_\alpha) = 0 \,\} \, .
\end{equation}
These sets $\mathcal{V}(P_\alpha)$ can then be compared with the sets $v_j$
obtained from the sampled points,
providing a map between sampled points $\underline X^{(0)}_j$ and ideals $P_\alpha$. 
The latter are candidate associated primes
appearing in the primary decomposition of $J$, and are taken to have the same dimension as $J$.
We verify these candidates by saturation,
\begin{equation}
J : P_\alpha^\infty =
\begin{cases}
J, & \text{if } P_\alpha \not\in \operatorname{Assoc}(J), \\[4pt]
\tilde J \supsetneq J, & \text{otherwise} .
\end{cases}
\end{equation}
In this way, the numerical sampling step guides the identification of components,
while saturation provides algebraic confirmation.

After saturation by all candidates $P_{\alpha_1}, \dots, P_{\alpha_m}$, we are left with either of two cases: 
\begin{equation}
J : P_{\alpha_1}^\infty \cdots : P_{\alpha_m}^\infty =
\begin{cases}
\langle 1 \rangle,
  & \text{if } \operatorname{Assoc}(J) \subseteq \{P_{\alpha_1},\dots,P_{\alpha_m}\} \,, \\[4pt]
\displaystyle\bigcap_{k:P_k \in \operatorname{Assoc}(J) \setminus \{P_{\alpha_1},\dots,P_{\alpha_m}\}} Q_k = \tilde J,
  & \text{otherwise} \,.
\end{cases}
\end{equation}
Here we use the fact that, in the cases considered in this work,
the primary components have uniform dimension, so that no additional
embedded components are removed by these saturations.
Like in previous computations \cite{DeLaurentis:2022otd, DeLaurentis:2025dxw}, 
we observe that at most one new primary component appears at a time. 
This is verified by explicitly testing the primaryness of the residual component,
using the test described in \cref{sec:primalitytest}.

\subsection{Lorentz-covariant generators via tensor ans\"atze}\label{sec:lorentz-cov-reconstruction}

After saturating by known primary components, we obtain an explicit list of
generators for the remaining ideal appearing in the primary decomposition,
\be\label{eq:jtilde_arbirary_generators}
    \tilde J = \langle f_1, \dots, f_L \rangle \, .
\ee
Our goal is to determine a minimal set of generators for $\tilde J$.
Moreover, since we are working in Lorentz-covariant rings (see
eq.~\eqref{eq:lorentz_covariant_ring}), we would like these generators to
have well-defined transformation properties under the Lorentz group.
In particular, we seek a generating set that transforms according to
eq.~\eqref{eq:Lorentz_scalar_function} or eq.~\eqref{eq:covariant_generator}.
Explicitly, we aim to write a set of generators of the form
\be\label{eq:ideal_with_covariant_generators}
    \tilde J =
    \langle
    a_1, \dots, a_{n_a},
    b_1^\alpha, \dots, b_{n_b}^\alpha,
    c_1^{\dot\alpha}, \dots, c_{n_c}^{\dot\alpha},
    d_1^{\alpha\dot\alpha}, \dots, d_{n_d}^{\alpha\dot\alpha},
    \dots
    \rangle \, ,
\ee
allowing for tensors carrying as many open spinor indices as required. 
Here and in the following, an ideal generated by covariant tensors
is understood as the ideal generated by all of their spinor components.
Since the generators $f_l$ are polynomials in the spinor components, $\lambda^{\alpha=1}_i,\lambda^{\alpha=2}_i,\tilde\lambda^{\dot\alpha=\dot 1}_i,\tilde\lambda^{\dot\alpha=\dot 2}_i$, one might
expect that each $f_l$ corresponds to a particular component of
one of the covariant generators appearing in
eq.~\eqref{eq:ideal_with_covariant_generators}.
However, after saturation one may encounter generators that mix invariant
and covariant contributions,
\begin{equation}
  f(\lambda,\tilde\lambda)
 =
 g(\lambda,\tilde\lambda)
 +
 h(\lambda,\tilde\lambda)\, ,
\end{equation}
where $g$ transforms covariantly while $h$ is Lorentz invariant, that is, schematically
\begin{equation}
 f(\Lambda \lambda,\, \tilde\Lambda \tilde\lambda)
 =
 \Lambda\tilde\Lambda\, g(\lambda,\tilde\lambda)
 +
 h(\lambda,\tilde\lambda)\, ,
\end{equation}
so that $f$ itself does not possess a definite transformation property. 
More generally, individual generators may contain contributions transforming
in several different representations of the Lorentz group. Concretely, this
corresponds to terms carrying different numbers of open spinor indices.

The method described in this section allows us to recover a set of covariant generators in the form of eq.~\eqref{eq:ideal_with_covariant_generators} from an arbitrary set of the form of eq.~\eqref{eq:jtilde_arbirary_generators} by exploiting the non-uniqueness of generating sets for ideals. In particular, we will fit ans\"atze for the covariant generators while sampling the mixed representation polynomials at phase-space points that isolate the various covariant contributions.

We begin by generalizing the spinor ansatz \cite{Laurentis:2019bjh, DeLaurentis:2022otd} to polynomials carrying open indices. The parameters defining the ansatz are unchanged, they are the \emph{mass dimension}, \(m\), and the \emph{phase weights}, \(\underline w =(w_1,\dots,w_n)\), of the target polynomial. For Lorentz-covariant quantities, the mass dimension $m$ can be half-integer (since spinors $\lambda^\alpha, \tilde\lambda^{\dot\alpha}$ have units of $\sqrt{E}$), whereas purely invariant objects must have integer $m$.  Therefore, a half-integer $m$ implies at least one open index (or really an odd number of open indices), while an integer $m$ implies an even number of open indices (incl.~no open index). For now, let us restrict our attention to the case of one open index, although a similar procedure could be generalized to an arbitrary number. 

We build the covariant ansatz, i.e.~a spanning set of monomials, as
\begin{equation}\label{eq:holomorphic-ansatz}
    \text{ansatz}(m, \underline w)
    =
    \bigcup_i
    \lambda^\alpha_i \,
    \text{ansatz}\!\left(
        m - \frac{1}{2},
        \underline w - \underline e_i
    \right),
    \qquad
    m \in \mathbb{N}_0 + \frac{1}{2} \, ,
\end{equation}
or, similarly, as
\begin{equation}\label{eq:antiholomorphic-ansatz}
    \text{ansatz}(m, \underline w)
    =
    \bigcup_i
    \tilde\lambda^{\dot\alpha}_i \,
    \text{ansatz}\!\left(
        m - \frac{1}{2},
        \underline w + \underline e_i
    \right),
    \qquad
    m \in \mathbb{N}_0 + \frac{1}{2} \, .
\end{equation}
Here $\underline e_i$ denotes the unit vector in phase-weight space
which shifts the weight of leg $i$ by one unit.
We omit potential combinations with more open indices,
and have adjusted the mass dimension and phase weights
on the RHS according to the factored spinor. The ans\"atze in the RHS are then standard invariant ones. Moreover, only one of eq.~\eqref{eq:holomorphic-ansatz} or eq.~\eqref{eq:antiholomorphic-ansatz} will ever be non-empty for a given $m$ and $\underline w$, because an invariant ansatz must satisfy
\begin{equation}
    \left( \frac{1}{2} \, \sum_j w_j \right) \; \text{mod} \;2 \, = \, m \; \text{mod} \; 2 \, . 
\end{equation}
This condition is necessary but not sufficient for the
existence of an invariant ansatz; further constraints follow
from the minimal mass dimension required to realise the specified
phase weights.
In practice, we construct the candidate monomial sets appearing
in eqs.~\eqref{eq:holomorphic-ansatz} and
\eqref{eq:antiholomorphic-ansatz} explicitly,
and remove redundancies from these unions by Gaussian elimination. For the present work, a single open index sufficed in all considered cases. However, we are aware of one example of primary decomposition where a primary (non-radical) ideal appears with 2 open indices, and then reduces to 1 open index upon taking the radical (i.e.~for the associated prime), see ref.~\cite[eq.~3.34]{Campbell:2022qpq}.

With a covariant ansatz at hand, we turn our attention to recovering a manifestly covariant form for the generators $f_1,\ldots,f_l$ of $\tilde J$, that is expressing eq.~\eqref{eq:jtilde_arbirary_generators} in the form of eq.~\eqref{eq:ideal_with_covariant_generators}. We do this through linear algebra over finite fields, by solving a linear system of equations of the form 
\begin{equation}\label{eq:ansatz-linear-system}
  A\,c = b \, , \qquad \text{with} \quad A_{ij} = M_j(\underline X^{(0)}_i) \quad \text{and} \quad b_i = f(\underline X^{(0)}_i) \, ,
\end{equation}
and then lifting the coefficient vector $c$ back to the rationals. Here $M_j$ denote the monomials in the ansatz and $\underline X^{(0)}_i$ random points in the appropriate (quotient) ring. While this approach works out of the box for some of the generators $f_l$, usually those with the lowest dimension $m$, it is insufficient in general. In fact, whenever a generator $f_l$ is not covariant or invariant such an ansatz would fail to span the required space. We can see how this comes to be in a simple example. Let $L=2$, and $\text{deg}(f_1) < \text{deg}(f_2)$. Then, $f_1$ must have good transformation properties under the Lorentz group automatically, since it is unique, while $f_2$ may be replaced by $\tilde f_2$, with 
\begin{equation}
    f_2 = \tilde f_2 + q f_1
\end{equation}
for any polynomial $q$, and still generate the same ideal. Crucially, $\tilde f_2$ and $q f_1$ need not live in the same representation of the Lorentz group. This suggests a simple resolution to our problem: instead of solving eq.~\eqref{eq:ansatz-linear-system} by sampling at random phase space points, if we sample the variety $V(\langle f_1 \rangle)$, then the sampled values of $f_2$ agree with those of $\tilde f_2$.
This allows us to reconstruct a covariant representative of $f_2$
modulo $\langle f_1\rangle$. In general, for an ideal generated by more polynomials, we may have to sample on varieties whose associated ideals have more than a single generator, but the general picture holds unchanged.

Following this procedure, we express all generators in the form of eq.~\eqref{eq:ideal_with_covariant_generators}. These are presented in \cref{sec:new-prim-dec} and \cref{5pt1m}.

\subsection{Primaries from associated primes}

The procedures described in \cref{sec:variety_sampling} allow one to identify candidate associated
primes $P_i$ appearing in the primary decomposition of an ideal
\begin{equation}
    I = \bigcap_{i=1}^{n_Q(I)} Q_i,
    \qquad
    P_i = \sqrt{Q_i}.
\end{equation}
However, knowing the associated prime $P_i$ does not, in general,
determine the corresponding primary component $Q_i$.
If the component is prime, then $Q_i=P_i$, but non-radical primary
components contain additional multiplicity information which is not
encoded in the associated prime $P_i=\sqrt{Q_i}$.

In the decompositions computed for previous amplitude applications,
this issue did not arise.
For the massless five-point finite remainders considered in
ref.~\cite{DeLaurentis:2022otd}, all primary components encountered
were prime.
In the later five-point one-mass decompositions of
ref.~\cite{DeLaurentis:2025dxw}, the only non-radical case involved
an ideal whose decomposition consisted of a single primary component.
Thus the ideal itself was the desired primary component, and no
separation of a non-radical primary from a larger intersection was
required.
In the present work, the inclusion of $\text{tr}_5$-dependent varieties
leads to non-radical primary components in non-trivial decompositions
with multiple associated primes.

The goal of this section, therefore, is the following.
Given an ideal $I$, together with an associated prime $P_i$ known to
occur in its primary decomposition, we want to identify a compact
set of generators for the corresponding primary component $Q_i$.
We describe two complementary strategies: an analytic one, based on
locating $Q_i$ within the filtration by powers of $P_i$, and a numerical
one, based on probing the order of vanishing near the variety $V(P_i)$.

\paragraph{Analytic approach.}
Let $Q$ be a $P$-primary component of $I$, so that
$\sqrt Q=P$.
Then $Q\subseteq P$.
Moreover, since every element of $P$ is nilpotent modulo $Q$,
there exists an integer $a$ such that
\begin{equation}
    P \supseteq Q \supseteq P^a \, .
\end{equation}
More generally, one may locate $Q$ between two levels of the
filtration by powers of $P$,
\begin{equation}\label{eq:P-power-filtration}
    P \supseteq P^2 \supseteq P^3 \supseteq \cdots ,
\end{equation}
namely
\begin{equation}\label{eq:primary-sandwich}
    P^b \supseteq Q \supseteq P^a \, ,
    \qquad b\leq a \, .
\end{equation}
The largest $b$ and smallest $a$ satisfying
\cref{eq:primary-sandwich} can be determined by direct inclusion tests. Note that it isn't necessarily required that the two filtration levels be adjacent.
The smallest admissible value of $a$ can also be read off from the
saturation index of $P$ in $I$, namely the first integer $a$ for which
\begin{equation}
    I:P^a = I:P^{a+1} \, ,
\end{equation}
provided that saturation by $P$ removes only the $P$-primary component.
While this filtration is straightforward to compute, its geometric
interpretation is not completely direct, since ordinary powers of $P$
need not be $P$-primary and may contain embedded components.

A more geometric interpretation is obtained by considering symbolic
powers.
We denote by $P^{\langle a\rangle}$ the $a$th symbolic power
of $P$, defined algebraically as the $P$-primary component of $P^a$.
The symbolic power $P^{\langle a\rangle}$ contains the polynomials
which vanish to order at least $a$ along $V(P)$, and in general one has
\begin{equation}
    P^{\langle a\rangle} \supseteq P^a \, .
\end{equation}
From this viewpoint, one can analogously locate $Q$ between two
levels of the symbolic-power filtration,
\begin{equation}
    P^{\langle b'\rangle} \supseteq Q
    \supseteq P^{\langle a'\rangle} \, ,
\end{equation}
for suitable integers $a'\ge b'$.\footnote{
The existence of such an $a'$ is non-trivial, but follows from a characterisation of symbolic powers by localisation.
}
These integers need not coincide with the integers $a,\,b$ appearing in
the ordinary-power filtration in eq.~\eqref{eq:primary-sandwich}.
In practice, symbolic powers are more expensive to compute, whereas
ordinary powers are straightforward to generate.
We therefore work with ordinary powers of $P$ and verify the required
containments directly.

The construction of $Q$ then proceeds from the inclusions in
eq.~\eqref{eq:primary-sandwich}.
We start with the generators of $P^a$, which are guaranteed to belong
to $Q$.
These generators need not generate all of $Q$.
We therefore iterate over the generators of the intermediate powers
$P^{a-1},P^{a-2},\dots,P^b$ and test, for each such generator $f$,
whether $f\in Q$.
Whenever this membership test succeeds, we add $f$ to the candidate
generating set.
Once this process terminates, we verify that the ideal generated by
the selected polynomials is equal to $Q$ and reduce the resulting
generating set to a minimal basis.

In principle, this procedure may miss generators which arise only as
non-trivial polynomial combinations of generators of the powers $P^i$,
rather than appearing directly among the chosen generators.\footnote{
For example, in $R=\mathbb{F}[x,y]$, let
$P=\langle x,y\rangle$ and
$Q=P^2+\langle x+y\rangle$.
Then $P\supseteq Q\supseteq P^2$, but the generator $x+y$ is not one
of the standard generators $x,y$ of $P$.
}
In the Lorentz-covariant applications considered here, this possibility
is strongly constrained by mass dimension and little-group weights:
candidate generators must be homogeneous in these gradings, so many
polynomial combinations that would be allowed in a generic polynomial ring
do not arise.
Moreover, some lower-filtration generators can be identified directly
from the generators of the original ideal $I$, even when they are
non-trivial polynomial combinations of generators of the powers $P^i$.
Nevertheless, if a more general situation were encountered, one could
first identify as many generators as possible through the filtration
procedure described above, and then reconstruct any remaining ones
using the ansatz-fitting strategy of
\cref{sec:lorentz-cov-reconstruction}.
For the examples considered in this work, see
\cref{sec:new-prim-dec}, this additional step was not required.

\paragraph{Numerical approach.}
The existence of a non-radical primary component $Q$, or equivalently
of a thickening of the associated variety $V(P)$, can also be probed
numerically.  This provides information about the position of $Q$ in
the filtration by powers of $P$, even though the identification of the
precise lower-filtration generators may still require analytic input.
The idea is to construct $p\kern0.2mm$-adic phase-space points which
approach $V(P)$, and to inspect the orders of vanishing of the
generators of the original ideal at such points.

Concretely, we work over $\mathbb{Q}_p$ and construct a point
$\underline X^{(\epsilon)}$ close to $V(P)$ such that, for a set of
generators
\begin{equation}
    P=\langle f_1,\dots,f_n\rangle ,
\end{equation}
one has
\begin{equation}
    f_i(\underline X^{(\epsilon)})=\mathcal{O}(p),
    \qquad i=1,\dots,n .
\end{equation}
Elements of $P^r$ then vanish at least as $\mathcal{O}(p^r)$
at such a point.  Thus, if some generators of
\begin{equation}
    I=\langle g_1,\dots,g_m\rangle
\end{equation}
are observed to vanish to higher than linear order at $\underline X^{(\epsilon)}$, this is a strong indication that the corresponding
primary component contains a thickening of $V(P)$.

This approach, however, involves subtleties which depend on how the
$p\kern0.2mm$-adic point approaching $V(P)$ is constructed.
In particular, the valuation of an individual polynomial at a chosen
point should not be identified directly with its filtration level in
the primary component.  A generator of $I$ may vanish to different
orders at different points approaching the same associated variety.
The numerical test should therefore be understood as a diagnostic for
the thickening of the component, warranting further analytic
investigation as described above, rather than as a complete alternative
way of determining the generating set.

\paragraph{Example: five-point massless ideal with $\text{tr}_5$.}
Consider the ideal
\begin{equation}
    I=\big\langle \langle12\rangle,\text{tr}_5(1,2,3,4)\big\rangle
    \subset R_5 \, ,
\end{equation}
for which we give a primary decomposition in
\cref{sec:res-five-point-massless}.
One of its associated primes is the all-angle-bracket ideal
\begin{equation}
    P_{\angle}
    =
    \big\langle
    \langle12\rangle,\langle13\rangle,\langle14\rangle,\langle15\rangle,
    \langle23\rangle,\langle24\rangle,\langle25\rangle,
    \langle34\rangle,\langle35\rangle,\langle45\rangle
    \big\rangle \, .
\end{equation}
The corresponding primary component is not simply $P_{\angle}$.
Instead, direct inclusion and saturation tests show that it lies
between $P_{\angle}$ and $P_{\angle}^2$, with saturation index $a=2$
in the sense of \cref{eq:primary-sandwich}.

The same behaviour is visible numerically.  Near $V(P_{\angle})$, all
angle brackets vanish to first order, while square brackets are
generically non-vanishing.  Using
\begin{equation}
    \text{tr}_5(1,2,3,4)
    =
    [12]\langle23\rangle[34]\langle41\rangle
    -
    \langle12\rangle[23]\langle34\rangle[41],
\end{equation}
we see that $\text{tr}_5$ has no generic linear term along this
branch.  Indeed, if all angle brackets are $\mathcal{O}(p)$, then
\begin{equation}
    \text{tr}_5(1,2,3,4)=\mathcal{O}(p^2).
\end{equation}
This provides a numerical indication of a quadratic thickening of the
angle-bracket component.

The full primary component is
\begin{equation}
    Q =
    P_{\angle}^2+\big\langle \langle12\rangle\big\rangle \, .
\end{equation}
The quadratic part $P_{\angle}^2$ reflects the fact that
$\text{tr}_5$ vanishes generically to second order along $V(P_{\angle})$, while
the additional generator $\langle12\rangle$ is inherited from the
original ideal $I$.  This illustrates the complementary roles of the
two methods: the $p\kern0.2mm$-adic diagnostic identifies the
thickening, while the analytic filtration determines the additional
generators needed to recover the full primary component.

\section{A test for prime and primary ideals}\label{sec:primalitytest}

It is often useful to validate the result of an involved computation by means
of a simpler one, even if the latter only yields a binary outcome (correct or
incorrect). In Ref.~\cite{DeLaurentis:2022otd}, a primality test was introduced
to verify the correctness of a primary decomposition for unmixed ideals.
However, as originally presented, this algorithm can itself be computationally
demanding, and may fail to terminate within a reasonable time for some of the
ideals encountered in subsequent applications.

In this section, we build on the algorithm of
Ref.~\cite{DeLaurentis:2022otd}, introducing two substantial improvements:
a semi-numerical strategy for computing the codimension of auxiliary ideals
arising at intermediate steps, and an extension of the primality criterion to
a full test of primaryness. The algorithm described here is implemented in
\textsc{syngular}~\cite{syngular},
see \texttt{ideal.test\_primality}.

Let us begin by briefly recalling the original algorithm. Given a tentative
primary decomposition, we write
\begin{equation}
    I = Q_1 \cap \dots \cap Q_n \, ,
\end{equation}
where the $Q_i$ are assumed to be primary,
with associated radicals $P_i$ assumed to be prime,
\begin{equation}
    \sqrt{Q_i} = P_i \, .
\end{equation}
The fundamental relation underlying the test is the decomposition
\cite{becker2012groebner}[Lemma~8.96]
\begin{equation}\label{eq:splittinglemma}
    J = (J + f^s) \cap J^{ec} \, ,
\end{equation}
where the extension and contraction are defined in
eqs.~\eqref{eq:extension} and~\eqref{eq:contraction},
the extended--contracted ideal $J^{ec}$ in
eq.~\eqref{eq:extension-contraction-through-saturation},
the polynomial $f$ in eq.~\eqref{eq:fpoly},
and $s$ denotes the saturation index of $f$ in $J$.
These operations are understood to be taken with respect to a maximally independent set $\underline Y$, such that $J^e$ is zero dimensional.

An ideal $J$, for instance taken to be one of the $P_i$, is then verified to be prime if the following two conditions hold:
\begin{enumerate}
    \item The extended--contracted ideal $J^{ec}$ is prime;
    \item The term $(J + f^s)$ is redundant in the intersection.
\end{enumerate}
The first condition is ensured by verifying that $J^{e}$ is maximal,
while the second follows from the inequality
\begin{equation}
    \dim(J + f^s) < \dim(J) \, ,
\end{equation}
under the assumption that $J$ is unmixed.

\subsection{Extension to primary ideals}

We now address the first of the two conditions above,
namely the requirement that the extended--contracted ideal $J^{ec}$ be prime.
The criterion of Ref.~\cite{DeLaurentis:2022otd}
can be improved in two directions.
First, it applies beyond the case in which the extended ideal $J^e$
is maximal.
Second, it can be used to determine not only whether an ideal is prime,
but also whether it is primary.

The key algebraic observation is that primaryness is preserved under
contraction. In particular, we have
\cite{becker2012groebner}[Lemma~8.97~(iii)]
\begin{equation}
    J^{e} \text{ primary in } \mathbb{F}(\underline Y)[\underline Z]
    \;\Longrightarrow\;
    J^{ec} \text{ primary in } \mathbb{F}[\underline X] \, .
\end{equation}
Thus, it is sufficient to determine whether the zero-dimensional
extension $J^e$ is primary.

In practice, rather than computing a primary decomposition of $J^e$
directly over $\mathbb{F}(\underline Y)$, which may be computationally
prohibitive, we specialise the independent variables to random values
$\underline Y\mapsto\underline Y^{(0)}$ in a large finite field.
We denote the resulting zero-dimensional ideal by $J^{(0)}$.
In the original formulation of the test in
Ref.~\cite{DeLaurentis:2022otd}, the sufficient condition that $J^e$
be maximal was already checked semi-numerically by looking for a
specialisation for which $J^{(0)}$ describes a single reduced point.
We generalise this step by computing the full primary decomposition of
$J^{(0)}$, for instance using the \texttt{primedecGTZ} algorithm
\cite{gianni1988grobner}.
If the decomposition contains a single primary component, this provides
evidence that $J^e$ is primary; if this component additionally coincides
with its radical, it provides evidence that $J^e$ is prime.
Since $J^{(0)}$ is zero-dimensional and the computation is performed
over a finite field, this step is inexpensive in practice.

Exceptional specialisations may, however, change the
primary-decomposition structure and therefore need not reflect the
properties of $J^e$.
This can lead to both false-negative and false-positive outcomes,
with rather different probabilities.
For the former, consider the prime ideal
\begin{equation}
    J=\langle x^2-y\rangle\subset\mathbb{F}[x,y] \, ,
\end{equation}
with $\underline Y=\{y\}$ and $\underline Z=\{x\}$.
Its extension $J^e\subset\mathbb{F}(y)[x]$ is also prime.
After specialising $y\mapsto a\in\mathbb{F}_p$, we have
\begin{equation}
    J^{(0)}=\langle x^2-a\rangle\subset\mathbb{F}_p[x] \, .
\end{equation}
For non-zero $a$, this ideal is prime when $a$ is a nonsquare and
non-primary when $a$ is a square.
Thus, approximately half of the specialisations give a negative result
even though $J^e$ is prime.
More generally, the probability of a false negative need not decrease
as the size of the finite field is increased.
A non-primary specialisation therefore cannot be taken as conclusive,
and the test is repeated over many independent choices of
$\underline Y^{(0)}$ until the required number of primary
specialisations is found or the prescribed number of attempts is
exhausted.

Conversely, an exceptional specialisation may cause distinct generic
components to merge.
For example, consider
\begin{equation}
    J=\langle x(x-y)\rangle\subset\mathbb{F}[x,y] \, .
\end{equation}
Its extension $J^e\subset\mathbb{F}(y)[x]$
is not primary, whereas the exceptional specialisation $y\mapsto0$
gives
\begin{equation}
    J^{(0)}=\langle x^2\rangle \, ,
\end{equation}
which is primary.
For uniformly sampled values in $\mathbb{F}_p$, this particular false
positive occurs with probability $1/p$.
More generally, such false positives arise only on exceptional
algebraic loci in the space of specialisations.
If such a locus is contained in the zero set of a non-zero polynomial
of degree $d$, its probability under a random specialisation over
$\mathbb{F}_p$ is at most of order $d/p$.
Requiring primaryness to be reproduced at $k$ independent
specialisations therefore suppresses the false-positive probability
as $(d/p)^k$, up to the combinatorial factor associated with allowing
a finite number of attempts.
In the new implementation, the number of required confirmations is
introduced as a parameter, with a default value of two.\footnote{For the chosen prime $p=536870909$, the largest prime below
$2^{29}$, even assuming an exceptional locus contained in the zero set
of a polynomial of degree $100$, the probability of obtaining two
false-positive confirmations among $100$ independent random
specialisations is estimated to be of order $10^{-10}$.}
When testing primality, the successful specialisations are additionally
required to agree on whether the unique primary component coincides
with its radical.

The resulting procedure is therefore asymmetric.
Negative specialisations are treated as inconclusive and the test is
repeated, suppressing false negatives as the number of attempts is
increased.
Once a positive specialisation is found, only a small number of
independent confirmations is required to rapidly suppress the probability
of an exceptional false positive.
Having established primaryness or primality of $J^e$ to the desired
statistical confidence, the corresponding property follows for
$J^{ec}$.
Once the term $(J+f^s)$ is shown to be redundant in the intersection,
the corresponding conclusion follows for $J$.

\subsection{Further semi-numerical improvements}\label{sec:seminumimprov}

We now address the second of the two conditions above,
namely the requirement that the term $(J + f^s)$ be redundant
in the intersection.
The computational complexity of the algorithm presented
in Ref.~\cite{DeLaurentis:2022otd}
is primarily driven by the polynomial $f$
appearing in the splitting lemma, eq.~\eqref{eq:splittinglemma},
which, we recall, is computed through eq.~\eqref{eq:fpoly}.
More precisely, it is governed by the complexity
of the irreducible factors of $f$.
Indeed, if $f = \prod_i g_i$ is the factorisation
of $f$ into irreducible factors, then
\cite{becker2012groebner}[Lemma~8.52] implies that
\begin{equation}
    \dim(J + \langle f \rangle)
    =
    \max_i \, \dim(J + \langle g_i \rangle) \, .
\end{equation}
In particular, verifying the dimension drop for $(J + \langle f \rangle)$
reduces to checking it for each irreducible factor $g_i$.
In practice, however, computing $\dim(J + \langle g_i \rangle)$
can remain computationally demanding when the factors $g_i$
are not simple.

The complexity of the factors $g_i$ depends on the choice
of maximal independent set $\underline Y$.
It was already observed in Ref.~\cite{DeLaurentis:2022otd}
that the freedom in the choice of $\underline Y$
can be exploited to minimise the complexity of $f$,
and hence that of the test itself.
In practice, however,
the function \texttt{indepSet} provided by \textsc{Singular}
often returns only a subset of the available maximal independent sets.
As a result, the choice of $\underline Y$ obtained in this way
need not be optimal, and may lead to polynomials $f$
with unnecessarily complicated factorisations.
This behaviour can be traced to the fact that
\texttt{indepSet} computes independent sets of the leading ideal,
i.e.\ the ideal generated by the leading monomials,
rather than of the ideal itself.

We improve upon this by exploiting numerical sampling of the variety.
By repeatedly generating generic points on $V(J)$
and inferring independent sets,
we obtain a more exhaustive set of candidates
than those accessible from the leading ideal alone.\footnote{
See Section~\ref{sec:variety_sampling}
for a detailed discussion of point generation and independent-set inference.
In practice, this is implemented using the numerical point-generation routine
\texttt{point\_on\_variety} described there,
with a dedicated ``guessing'' mode for independent sets
(\texttt{indepSet=\textquotesingle force guess\textquotesingle}).
The sampling is parallelised to efficiently explore
a sufficiently large number of generic points on the variety.
}
In practice, this substantially increases the likelihood
of identifying an independent set
for which the resulting polynomial $f$ has low degree
and simple irreducible factors $g_i$.
The test is extended to accept an independent set
as input, allowing favourable choices identified during sampling
to be cached and reused.
This enables subsequent runs to bypass the search step
and significantly reduces the computational cost of reproducing the result.

Even with the enlarged search space for favourable independent sets,
computing the dimension of the auxiliary ideals $(J + \langle g_i \rangle)$
through standard symbolic methods, such as those implemented in
\textsc{Singular}, can remain computationally expensive.
To address this, we apply the same semi-numerical strategy
described in Sec.~\ref{sec:variety_sampling}.
In particular, we generate random numerical points on
$V(J + \langle g_i \rangle)$ to infer an independent set, 
from which the dimension of the corresponding ideal is inferred.
This mirrors the procedure used for $J$ itself,
and allows the dimension of $(J + \langle g_i \rangle)$
to be determined efficiently in practice.

\section{New primary decompositions}\label{sec:new-prim-dec}

We now present the new codimension-two primary decompositions associated
with the denominator factors discussed in \cref{sec:irred_denoms}.
We give only one representative for ideals related by permutations of
kinematically equivalent external legs or by parity conjugation, since
the primary decomposition of any such related ideal is obtained by
applying the same transformation to each of its primary components.
We denote by $P(i_1i_2\ldots i_n)$ the ideal obtained from $P$ by the
relabelling $j\mapsto i_j$, and by $\bar P$ its parity conjugate, defined
by $\lambda_i^\alpha \leftrightarrow \tilde\lambda_i^{\dot\alpha}$.

\subsection{Five-point massless}\label{sec:res-five-point-massless}

Working in the Lorentz-covariant ring $R_5$ introduced in
\cref{sec:kinemrev}, the inclusion of $\trFive$, as discussed in
\cref{sec:irred_denoms}, gives rise to two inequivalent codimension-two
ideals not covered in ref.~\cite{DeLaurentis:2022otd}. We choose the
representatives
\begin{equation}
    \big\langle \langle 12 \rangle, \; \trFive(1,2,3,4) \big\rangle_{R_5}
    \qquad\text{and}\qquad
    \big\langle \langle 1|2+3|1 ], \; \trFive(1,2,3,4) \big\rangle_{R_5}.
\end{equation}
Most of the associated primes appearing in their primary decompositions
were already identified in ref.~\cite{DeLaurentis:2022otd}, where they were
denoted by

\begin{equation}
P_1 = \big\langle \langle12\rangle, \langle13\rangle,  \langle23\rangle , [45] \big\rangle_{R_5} 
\end{equation}
\begin{equation}
P_2 = \big\langle |1\rangle \big\rangle_{R_5} 
\end{equation}
\begin{equation}
P_3 = \big\langle 
\langle 12\rangle, \langle 23\rangle, \langle 34\rangle, \langle 45\rangle, \langle 15\rangle, \langle 13\rangle, \langle 14\rangle, \langle 24\rangle, \langle 25\rangle, \langle 35\rangle
\big\rangle_{R_5} 
\end{equation}
\begin{equation}
P_5 = \big\langle \langle12\rangle, [12] \big\rangle_{R_5} 
\end{equation}
In terms of these, the two new primary decompositions read
\begin{equation}\label{eq:primdecangle12tr5}
\begin{aligned}
\langle \langle 12 \rangle, \;\mathrm{tr}_5(1,2,3,4) \rangle_{R_5} \;=\;\,
& P_1 \cap P_1(12543) \cap P_1 (12453) \cap \bar{P}_1(34512) \; \cap \\
& \quad P_2 \cap P_2(21543) \cap \big(P_3^2 + \langle \langle12\rangle\rangle\big) \cap P_5
\end{aligned}
\end{equation}
where we have made explicit the non-radical primary component associated with
the prime $P_3$.
The second new primary decomposition contains two non-radical components,
with associated primes $P_3$ and $\bar P_3$, respectively, as well as a
new prime ideal $P_{11}$,
\begin{equation}
\begin{aligned}
\langle \langle 1|2+3|1], \;\mathrm{tr}_5(1,2,3,4) \rangle_{R_5} \;=\;\, & P_1 \cap\bar{P}_1 \cap {P}_1(15432) \cap \bar{P}_1(15432) \; \cap P_2 \cap\bar{P}_2 \; \cap\\ 
& \quad (P_3 ^ 2 + \big\langle \langle 1|2+3|1]\big\rangle) \cap (\bar{P}_3 ^ 2 + \big\langle \langle 1|2+3|1]\big\rangle) \cap P_{11} \, .
\end{aligned}
\end{equation}
The new prime $P_{11}$ is given by
\begin{equation}
\label{eq:five-point-prime-11}
\begin{aligned}
P_{11} = \Big\langle\,
&\langle 1|(2+3)|1], \mathrm{tr}_5(1,2,3,4), \\
&\langle 13\rangle \langle 14\rangle \langle 25\rangle [23]
 + \langle 12\rangle \langle 15\rangle \langle 45\rangle [25],
 \quad \lambda_i^\alpha \leftrightarrow \tilde\lambda_i^{\dot\alpha}, \\
&\langle 13\rangle \langle 15\rangle \langle 23\rangle [35]
 - \langle 12\rangle \langle 14\rangle \langle 35\rangle [45],
 \quad \lambda_i^\alpha \leftrightarrow \tilde\lambda_i^{\dot\alpha}, \\
&\langle 23\rangle^2 [24][35]
 - \langle 24\rangle \langle 35\rangle [45]^2,
 \quad \lambda_i^\alpha \leftrightarrow \tilde\lambda_i^{\dot\alpha}, \\
&|2]\langle 12\rangle^2 \langle 15\rangle \langle 34\rangle
 + |3]\langle 13\rangle^2 \langle 15\rangle \langle 24\rangle 
 - |4]\langle 12\rangle \langle 13\rangle \langle 14\rangle \langle 45\rangle,
 \quad \lambda_i^\alpha \leftrightarrow \tilde\lambda_i^{\dot\alpha}
\,\Big\rangle_{R_5}.
\end{aligned}
\end{equation}
The generators of $P_{11}$ include several Lorentz-covariant polynomials
with open spinor indices. These were reconstructed using the procedure
described in \cref{sec:lorentz-cov-reconstruction}. For the final generator,
the polynomial obtained from \textsc{Singular} contained contributions
proportional to $\langle1|2+3|1]$. We therefore performed the
reconstruction on the variety
$V\Big(\big\langle\langle1|2+3|1]\big\rangle\Big)$, where these contributions vanish.
The reconstruction of this final generator is given explicitly in
\cref{app:covariant_ansatz}.

As a representative illustration of the computational improvements described in
\cref{sec:primalitytest}, the primality test for $P_{11}$ involves 88 factors
of the $f$-polynomial for the most favourable independent set found.
Using symbolic dimension computations, the test does not terminate within a
practical timescale. After approximately 20 minutes, it reaches only the $35^{\text{th}}$ factor in a list ordered by increasing complexity. The dimension computation for this factor alone exceeds a $600\,$s timeout, while several of the remaining factors are considerably more complicated. By contrast, using the
semi-numerical dimension computation described in
\cref{sec:seminumimprov}, the complete test takes approximately
8 minutes. This includes the generation of 100 numerical points on the variety,
used to infer additional candidate independent sets beyond those returned by
\textsc{Singular}; in this case, none improves on the most favourable set already
identified. Reusing that independent set, which yields the 88 factors quoted
above, the runtime is reduced to approximately 3.5 minutes.

\subsection{Five-point one-mass}
\label{sec:res-five-point-onemass}

We next consider five-point one-mass kinematics, working in the
Lorentz-covariant ring $R_6$ through the massless embedding described in
\cref{sec:irred_denoms}. Ref.~\cite{DeLaurentis:2025dxw} determined the
codimension-two primary decompositions constructed from the denominator
factors appearing at one loop, together with one selected decomposition
involving the additional denominator type appearing at two loops. Here we
complete the analysis of the latter.

We construct all denominator-pair ideals for which one of the two generators
is the new two-loop irreducible denominator factor, namely the final
representative in \cref{eq:five-point-one-mass-invariants}. Retaining one
representative under permutations of the four massless external legs and
parity conjugation, following the conventions introduced above, yields 27
inequivalent codimension-two ideals, denoted $I_1,\ldots,I_{27}$ in
\cref{5pt1mappendix}. We do not consider pairs in which both generators are
of the new two-loop type. Based on the examples studied so far, the LCD
numerators are expected to belong to the ideal generated by any pair of such
factors appearing in the LCD
\cite{DeLaurentis:2025dxw,DeLaurentis:2026brm}. Assuming this pattern holds,
primary decompositions of these ideals would provide little additional
information relevant to partial-fraction decompositions, while potentially
requiring considerably more involved computations.

The complete primary decompositions of the 27 ideals are given in
\cref{5pt1mappendix}. They contain both associated primes already encountered
in Ref.~\cite{DeLaurentis:2025dxw} and new ones. The previously known primes
that occur in the present decompositions are
\begin{equation}
    P_1,\;P_2,\;P_5,\;P_6,\;P_7,\;P_8,\;P_{10},\;P_{11},\;P_{19}
    \subset R_6\,,
\end{equation}
whose explicit forms are recalled in
\cref{eq:P1,eq:P2,eq:P5,eq:P6,eq:P7,eq:P8,eq:P10,eq:P11,eq:P19}.
These include familiar loci such as the holomorphic collinear configuration
described by
$P_1=\langle\langle12\rangle,\langle13\rangle,\langle23\rangle\rangle$, and the soft configuration described by
$P_2=\langle |1\rangle\rangle$.

In addition, the decompositions contain 13 new non-trivial associated primes, 
\begin{equation}
P_{57},\; P_{58},\; P_{59},\; P_{60},\; P_{61}, \;P_{62},\; P_{63},\; P_{64},\; P_{65},\; P_{66},\; P_{67},\; P_{68},\; P_{69} \subset R_6\,,
\end{equation}
whose explicit Lorentz-covariant generators are given in
\cref{5pt1mappendix}. There are also 4 further cases,
\begin{equation}
I_{12} = P_{70},\;I_{14} = P_{71},\;I_{25} = P_{72},\; I_{26} = P_{73},
\end{equation}
in which the denominator-pair ideal is itself prime. These new associated
primes arise from intersections involving the new two-loop denominator
structure and were not present in the one-loop analysis of
Ref.~\cite{DeLaurentis:2025dxw}.

The decompositions were obtained using the methods described in
\cref{sec:computing-primary-decompositions} and subjected to several
independent checks. For each ideal $I$, all primary components found in the decomposition are prime, so we explicitly verify the equality
$I=\bigcap_i P_i$. 
We continue to find at most one previously unknown component in each
decomposition. As an independent check, we therefore remove all previously
known components by successive saturations of $I$ with their associated
primes and verify that the remaining ideal coincides with the newly
identified prime. Finally, each new ideal labelled as $P_i$ is independently verified to be
prime using the test of \cref{sec:primalitytest}. For these tests, we retain
the maximal independent set yielding the simplest $f$-polynomial found,
allowing the results to be re-verified efficiently. We provide these tests as an ancillary file, \texttt{test\_primary\_decompositions.py}.

\section{Conclusions}
\label{sec:conclusions}

In this work, we have developed a more systematic approach to
computing primary decompositions in the Lorentz-covariant quotient
rings relevant to scattering amplitudes, extending the strategy of
Ref.~\cite{DeLaurentis:2022otd}. Our main motivation is the
singularity structure of rational amplitude coefficients. Individual
denominator factors define codimension-one varieties closely related
to Landau singular loci. These varieties are generally irreducible
beyond four-point massless kinematics, while their pairwise
intersections are often reducible. Primary decomposition resolves
these intersections into their irreducible branches while retaining
the multiplicity information encoded by non-radical components. We
have addressed several practical obstacles that arise in carrying out
these computations: the efficient determination of independent sets
and variety dimensions, the reconstruction of manifestly
Lorentz-covariant generators, the recovery of non-radical primary
components from their associated primes, and the verification of prime
and primary ideals. The generic algebraic functionality developed here
is implemented in \textsc{syngular}~\cite{syngular}, while its
application to the Lorentz-covariant kinematic rings relevant to
scattering amplitudes is implemented in
\textsc{lips}~\cite{DeLaurentis:2023qhd}. The ansatz-fitting routines
used for the reconstruction of Lorentz-covariant generators will be
made available in a future release of \textsc{antares}
\cite{giuseppe_de_laurentis_2026_18894183}.

We applied these methods first to five-point massless kinematics,
described by the quotient ring $R_5$, where the inclusion of the
parity-odd invariant $\trFive$ completes the codimension-two analysis
of Ref.~\cite{DeLaurentis:2022otd}. The resulting decompositions
contain non-radical primary components in non-trivial intersections
with several associated primes, as well as the new prime
$P_{11}\subset R_5$. We then considered five-point one-mass kinematics
through its embedding into massless six-point kinematics, working in
the quotient ring $R_6$, and completed the analysis of
denominator-pair ideals involving the additional irreducible
denominator structure appearing at two loops. This gives 27
inequivalent codimension-two ideals and introduces 17 new associated
primes beyond those already present in the analysis of
Ref.~\cite{DeLaurentis:2025dxw}. These decompositions provide the
algebraic information required to study multivariate partial-fraction
representations.

From a broader computational-algebraic perspective, the determination
of dimension and maximal independent sets from Gr\"obner bases is a
classical problem~\cite{DBLP:journals/jsc/KredelW88}, while the
generation of generic points on positive-dimensional varieties through
linear slicing has been studied in numerical algebraic
geometry~\cite{DBLP:journals/jc/SommeseV00} and, over finite fields,
is implemented for example in the \textsc{RandomPoints} package for
\textsc{Macaulay2}~\cite{RandomPointsSource,RandomPointsArticle,M2}. The
procedure used here instead specialises subsets of coordinate
variables chosen through independent sets. Once a suitable independent
set has been found, generic values are assigned to its variables and
the remaining problem is reduced to a zero-dimensional polynomial
system. This makes the same construction useful both for repeated
point generation and for the semi-numerical determination of
dimensions and independent sets that underlies the improved primality
test we developed. The implementation supports complex floating-point,
finite-field and $p\kern0.1mm$-adic arithmetic. The present
zero-dimensional solver relies on \textsc{Singular}; a natural
computational improvement would be to interface this step with
specialised solvers such as \textsc{msolve}~\cite{msolve}, which
provides highly optimised Gröbner-basis and polynomial-system-solving
algorithms over prime fields.

A natural next application is massless six-point kinematics. The
two-loop six-point alphabet of Ref.~\cite{Abreu:2024fei} provides a
natural set of candidate singular structures, although not all of
these are expected to appear as new denominator factors in the
rational coefficients. Once the relevant denominator factors are
identified, computing their codimension-two primary decompositions
would provide a first application of the present methods at genuine
six-point complexity. More generally, it would be valuable to
understand how primary decompositions at lower multiplicity are
embedded into those at higher multiplicity, and under what conditions
the resulting families of associated primes eventually stabilise. The
developments presented here provide a more concrete route towards
incorporating this geometric information directly into finite-field
reconstruction and multivariate partial-fraction algorithms. Combining
all of these ingredients into a fully automated reconstruction
strategy remains non-trivial, but would offer a systematic way of
reducing the analytic complexity of higher-loop and
higher-multiplicity scattering amplitudes.

\section*{Acknowledgments}
G.D.L.'s work is supported in part by the U.K.\ Royal Society through
Grant URF\textbackslash R1\textbackslash 20109. We thank Einan Gardi for comments on the manuscript.

\pagebreak

\begin{appendix}

\section{Example of covariant ansatz}\label{app:covariant_ansatz}
Using the component notation
\begin{equation}
\lambda_i = \begin{pmatrix} a_i \\ b_i \end{pmatrix} \quad
\tilde\lambda_i = \begin{pmatrix} c_i & d_i \end{pmatrix}
\end{equation}
an example of what a polynomial obtained from \textsc{Singular} may look like is 
\begin{align*}
f = \; & b_{1}^{4} c_{1} a_{2} a_{3} a_{4} a_{5}
- a_{1} b_{1}^{3} c_{1} b_{2} a_{3} a_{4} a_{5}
+ b_{1}^{3} a_{2} b_{2} c_{2} a_{3} a_{4} a_{5} \\
& - a_{1} b_{1}^{2} b_{2}^{2} c_{2} a_{3} a_{4} a_{5}
- a_{1} b_{1}^{3} c_{1} a_{2} b_{3} a_{4} a_{5}
+ a_{1}^{2} b_{1}^{2} c_{1} b_{2} b_{3} a_{4} a_{5} \\
& + b_{1}^{3} a_{2}^{2} c_{2} b_{3} a_{4} a_{5}
- 3 a_{1} b_{1}^{2} a_{2} b_{2} c_{2} b_{3} a_{4} a_{5}
+ 2 a_{1}^{2} b_{1} b_{2}^{2} c_{2} b_{3} a_{4} a_{5} \\
& + b_{1}^{3} b_{2} a_{3}^{2} c_{3} a_{4} a_{5}
+ b_{1}^{3} a_{2} a_{3} b_{3} c_{3} a_{4} a_{5}
- 3 a_{1} b_{1}^{2} b_{2} a_{3} b_{3} c_{3} a_{4} a_{5} \\
& - a_{1} b_{1}^{2} a_{2} b_{3}^{2} c_{3} a_{4} a_{5}
+ 2 a_{1}^{2} b_{1} b_{2} b_{3}^{2} c_{3} a_{4} a_{5}
- a_{1} b_{1}^{3} c_{1} a_{2} a_{3} b_{4} a_{5} \\
& + a_{1}^{2} b_{1}^{2} c_{1} b_{2} a_{3} b_{4} a_{5}
- b_{1}^{3} a_{2}^{2} c_{2} a_{3} b_{4} a_{5}
+ a_{1} b_{1}^{2} a_{2} b_{2} c_{2} a_{3} b_{4} a_{5} \\
& + a_{1}^{2} b_{1}^{2} c_{1} a_{2} b_{3} b_{4} a_{5}
- a_{1}^{3} b_{1} c_{1} b_{2} b_{3} b_{4} a_{5}
+ a_{1}^{2} b_{1} a_{2} b_{2} c_{2} b_{3} b_{4} a_{5} \\
& - a_{1}^{3} b_{2}^{2} c_{2} b_{3} b_{4} a_{5}
- b_{1}^{3} a_{2} a_{3}^{2} c_{3} b_{4} a_{5}
+ a_{1} b_{1}^{2} a_{2} a_{3} b_{3} c_{3} b_{4} a_{5} \\
& + a_{1}^{2} b_{1} b_{2} a_{3} b_{3} c_{3} b_{4} a_{5}
- a_{1}^{3} b_{2} b_{3}^{2} c_{3} b_{4} a_{5}
- a_{1} b_{1}^{3} c_{1} a_{2} a_{3} a_{4} b_{5} \\
& + a_{1}^{2} b_{1}^{2} c_{1} b_{2} a_{3} a_{4} b_{5}
- b_{1}^{3} a_{2}^{2} c_{2} a_{3} a_{4} b_{5}
+ a_{1} b_{1}^{2} a_{2} b_{2} c_{2} a_{3} a_{4} b_{5} \\
& + a_{1}^{2} b_{1}^{2} c_{1} a_{2} b_{3} a_{4} b_{5}
- a_{1}^{3} b_{1} c_{1} b_{2} b_{3} a_{4} b_{5}
+ a_{1}^{2} b_{1} a_{2} b_{2} c_{2} b_{3} a_{4} b_{5} \\
& - a_{1}^{3} b_{2}^{2} c_{2} b_{3} a_{4} b_{5}
- b_{1}^{3} a_{2} a_{3}^{2} c_{3} a_{4} b_{5}
+ a_{1} b_{1}^{2} a_{2} a_{3} b_{3} c_{3} a_{4} b_{5} \\
& + a_{1}^{2} b_{1} b_{2} a_{3} b_{3} c_{3} a_{4} b_{5}
- a_{1}^{3} b_{2} b_{3}^{2} c_{3} a_{4} b_{5}
+ a_{1}^{2} b_{1}^{2} c_{1} a_{2} a_{3} b_{4} b_{5} \\
& - a_{1}^{3} b_{1} c_{1} b_{2} a_{3} b_{4} b_{5}
+ 2 a_{1} b_{1}^{2} a_{2}^{2} c_{2} a_{3} b_{4} b_{5}
- 3 a_{1}^{2} b_{1} a_{2} b_{2} c_{2} a_{3} b_{4} b_{5} \\
& + a_{1}^{3} b_{2}^{2} c_{2} a_{3} b_{4} b_{5}
- a_{1}^{3} b_{1} c_{1} a_{2} b_{3} b_{4} b_{5}
+ a_{1}^{4} c_{1} b_{2} b_{3} b_{4} b_{5} \\
& - a_{1}^{2} b_{1} a_{2}^{2} c_{2} b_{3} b_{4} b_{5}
+ a_{1}^{3} a_{2} b_{2} c_{2} b_{3} b_{4} b_{5}
+ 2 a_{1} b_{1}^{2} a_{2} a_{3}^{2} c_{3} b_{4} b_{5} \\
& - a_{1}^{2} b_{1} b_{2} a_{3}^{2} c_{3} b_{4} b_{5}
- 3 a_{1}^{2} b_{1} a_{2} a_{3} b_{3} c_{3} b_{4} b_{5}
+ a_{1}^{3} b_{2} a_{3} b_{3} c_{3} b_{4} b_{5} \\
& + a_{1}^{3} a_{2} b_{3}^{2} c_{3} b_{4} b_{5}
\end{align*}
which corresponds to the final covariant generator recovered for
$P_{11}$. The mass dimension and phase weights are
\begin{equation}
m = 4.5,\quad \underline w = [3, 1, 1, 1, 1] \, . 
\end{equation}
Since the mass dimension is half-integer, the reconstructed covariant
carries an open spinor index. An ansatz satisfying these constraints
is
\begin{equation}
\begin{bmatrix}
|1] & \langle 1|2 \rangle & \langle 1|3 \rangle & \langle 1|4 \rangle & \langle 1|5 \rangle \\
|2] & \langle 1|2 \rangle & \langle 1|2 \rangle & \langle 1|4 \rangle & \langle 3|5 \rangle \\
|2] & \langle 1|2 \rangle & \langle 1|2 \rangle & \langle 1|3 \rangle & \langle 4|5 \rangle \\
|2] & \langle 1|2 \rangle & \langle 1|3 \rangle & \langle 1|4 \rangle & \langle 2|5 \rangle \\
|3] & \langle 1|3 \rangle & \langle 1|3 \rangle & \langle 1|4 \rangle & \langle 2|5 \rangle \\
|3] & \langle 1|2 \rangle & \langle 1|3 \rangle & \langle 1|3 \rangle & \langle 4|5 \rangle \\
|3] & \langle 1|2 \rangle & \langle 1|3 \rangle & \langle 1|4 \rangle & \langle 3|5 \rangle \\
|4] & \langle 1|2 \rangle & \langle 1|4 \rangle & \langle 1|4 \rangle & \langle 3|5 \rangle \\
|4] & \langle 1|2 \rangle & \langle 1|3 \rangle & \langle 1|4 \rangle & \langle 4|5 \rangle \\
|4] & \langle 1|3 \rangle & \langle 1|4 \rangle & \langle 1|4 \rangle & \langle 2|5 \rangle \\
|5] & \langle 1|3 \rangle & \langle 1|4 \rangle & \langle 1|5 \rangle & \langle 2|5 \rangle \\
|5] & \langle 1|2 \rangle & \langle 1|3 \rangle & \langle 1|5 \rangle & \langle 4|5 \rangle \\
|5] & \langle 1|2 \rangle & \langle 1|4 \rangle & \langle 1|5 \rangle & \langle 3|5 \rangle
\end{bmatrix}
\end{equation}
where each row corresponds to a monomial in the ansatz.

We denote by $M_j$, where $j=1,\ldots,13$, the covariant monomials
obtained by multiplying the entries in the $j$th row of the ansatz
above. Following \cref{eq:ansatz-linear-system}, the matrix $A$ and
vector $b$ are constructed by evaluating the ansatz monomials and the
polynomial component $f$ at the same phase-space points:
\begin{equation}
    A_{ij} = M_j\big(\underline X_i^{(0)}\big),
    \qquad
    b_i = f\big(\underline X_i^{(0)}\big).
\end{equation}
Since $M_j$ carries one open-square index spinor, the component corresponding to polynomial $f$ is selected for every entry of $A$. 

For example, the first two columns of $A$ are given by
\begin{align}
    A_{i1}
    &=
    \left.
    |1]\langle 1|2\rangle
        \langle 1|3\rangle
        \langle 1|4\rangle
        \langle 1|5\rangle
    \right|_{\underline X=\underline X_i^{(0)}},
    \\
    A_{i2}
    &=
    \left.
    |2]\langle 1|2\rangle^2
        \langle 1|4\rangle
        \langle 3|5\rangle
    \right|_{\underline X=\underline X_i^{(0)}}.
\end{align}
Thus, the $i$th row of $A$ contains the values of the ansatz monomials
at $X_i^{(0)}$, while the $j$th column contains the evaluations of
$M_j$. The vector $b$ contains the corresponding values of $f$.  We
solve $A c=b$ by using Gaussian elimination and the reconstruction
gives
\begin{equation}
\mathbf{c} =
\begin{pmatrix}
-1, -1, 1, -1, -1, 1, -1, 0, 0, 0, 0, 0, 0
\end{pmatrix} ^{T}
\in\mathbb{Q}^{13}.
\end{equation}
The entries of $c$ are the coefficients of the monomials in the order
they appear in the ansatz, and substituting the coefficients back into
the ansatz reproduces the polynomial component $f$ in the following
covariant form:

\begin{equation}
\begin{aligned}
f &= -\,|1]\langle 1|2\rangle \langle 1|3\rangle \langle 1|4\rangle \langle 1|5\rangle  -\,|2]\langle 1|2\rangle \langle 1|2\rangle \langle 1|4\rangle \langle 3|5\rangle \\
& +\,|2]\langle 1|2\rangle \langle 1|2\rangle \langle 1|3\rangle \langle 4|5\rangle -\,|2]\langle 1|2\rangle \langle 1|3\rangle \langle 1|4\rangle \langle 2|5\rangle \\
& -\,|3]\langle 1|3\rangle \langle 1|3\rangle \langle 1|4\rangle \langle 2|5\rangle +\,|3]\langle 1|2\rangle \langle 1|3\rangle \langle 1|3\rangle \langle 4|5\rangle \\
& -\,|3]\langle 1|2\rangle \langle 1|3\rangle \langle 1|4\rangle \langle 3|5\rangle.
\end{aligned}
\end{equation}
Through the Schouten identity and momentum conservation, this
simplifies, up to an overall sign, to the last element of $P_{11}$ at
\cref{eq:five-point-prime-11},
\begin{equation}
\begin{aligned}
|2]\langle 1|2\rangle^2 \langle 1|5\rangle \langle 3|4\rangle
 + |3]\langle 1|3\rangle^2 \langle 1|5\rangle \langle 2|4\rangle 
 - |4]\langle 1|2\rangle \langle 1|3\rangle \langle 1|4\rangle \langle 4|5\rangle \, .
 \end{aligned}
\end{equation}

\section{Five-Point One-Mass Primary Decompositions}\label{5pt1m}\label{5pt1mappendix}
We find the following decompositions for five-point one-mass
kinematics with the new two-loop denominator factor:
\allowdisplaybreaks
\begin{align}
I_1 &= \Big\langle \langle 1|2\rangle,  \langle 2|3|5+6|1|2]-\langle 3|4|5+6|1|3]  \Big\rangle_{R_6} \nonumber \\
&= P_{11}(124356) \cap P_6 \cap P_2 \cap P_1 , \\[6pt]
I_2 &= \Big\langle \langle 1|2\rangle,  \langle 3|2|5+6|1|3]-\langle 2|4|5+6|1|2]  \Big\rangle_{R_6} \nonumber \\
&= P_7 \cap P_5(213456) \cap P_2(213456) \cap P_2 \cap P_1 , \\[6pt]
I_3 &= \Big\langle \langle 1|2\rangle,  \langle 3|4|5+6|1|3]-\langle 4|2|5+6|1|4]  \Big\rangle_{R_6} \nonumber \\
&= P_{57} \cap P_2 , \\[6pt]
I_4 &= \Big\langle \langle 1|2\rangle,  \langle 1|2|5+6|3|1]-\langle 2|4|5+6|3|2]  \Big\rangle_{R_6} \nonumber \\
&= P_8(124356) \cap P_5(213456) \cap P_2(213456) \cap P_1(124356) \cap P_1 , \\[6pt]
I_5 &= \Big\langle \langle 1|2\rangle,  \langle 1|4|5+6|3|1]-\langle 4|2|5+6|3|4]  \Big\rangle_{R_6} \nonumber \\
&= P_{58} \cap P_1(124356) , \\[6pt]
I_6 &= \Big\langle \langle 1|2\rangle,  \langle 4|1|5+6|3|4]-\langle 1|2|5+6|3|1]  \Big\rangle_{R_6} \nonumber \\
&= P_{11} \cap P_6 \cap P_2 \cap P_1(124356) , \\[6pt]
I_7 &= \Big\langle \langle 1|2+3|1],  \langle 2|3|5+6|1|2]-\langle 3|4|5+6|1|3]  \Big\rangle_{R_6} \nonumber \\
&= P_{59} \cap \bar{P}_2 \cap P_2 \cap \bar{P}_1 \cap P_1 , \\[6pt]
I_8 &= \Big\langle \langle 1|2+3|1],  \langle 2|4|5+6|1|2]-\langle 4|3|5+6|1|4]  \Big\rangle_{R_6} \nonumber \\
&= P_{19}(132456) \cap \bar{P}_7(142356) \cap P_7(142356) \cap \bar{P}_2 \cap P_2 , \\[6pt]
I_9 &= \Big\langle \langle 1|2+3|1],  \langle 4|2|5+6|1|4]-\langle 2|3|5+6|1|2]  \Big\rangle_{R_6} \nonumber \\
&= P_{60} \cap \bar{P}_2 \cap P_2 , \\[6pt]
I_{10} &= \Big\langle \langle 1|2+3|1],  \langle 1|3|5+6|2|1]-\langle 3|4|5+6|2|3]  \Big\rangle_{R_6} \nonumber \\
&= P_{61} \cap \bar{P}_1 \cap P_1 , \\[6pt]
I_{11} &= \Big\langle \langle 1|2+3|1],  \langle 3|1|5+6|2|3]-\langle 1|4|5+6|2|1]  \Big\rangle_{R_6} \nonumber \\
&= \bar{P}_{10}(561234) \cap P_{10}(561234) \cap P_5(132456) \cap P_5 \\
&\quad \cap \bar{P}_2 \cap P_2 \cap \bar{P}_1 \cap P_1 , \\[6pt]
I_{12} &= \Big\langle \langle 1|2+3|1],  \langle 3|4|5+6|2|3]-\langle 4|1|5+6|2|4]  \Big\rangle_{R_6} \\
&= P_{70}, \\[6pt]
I_{13} &= \Big\langle \langle 1|2+3|1],  \langle 4|1|5+6|2|4]-\langle 1|3|5+6|2|1]  \Big\rangle_{R_6} \\
&= P_{62} \cap P_2 \cap \bar{P}_2 , \\[6pt]
I_{14} &= \Big\langle \langle 1|2+3|1],  \langle 4|3|5+6|2|4]-\langle 3|1|5+6|2|3]  \Big\rangle_{R_6} \nonumber \\
&= P_{71}, \\[6pt]
I_{15} &= \Big\langle \langle 1|2+3|1],  \langle 1|2|5+6|4|1]-\langle 2|3|5+6|4|2]  \Big\rangle_{R_6} \nonumber \\
&= P_{63} \cap \bar{P}_1 \cap P_1 , \\[6pt]
I_{16} &= \Big\langle \langle 1|2+3|1],  \langle 2|1|5+6|4|2]-\langle 1|3|5+6|4|1]  \Big\rangle_{R_6} \nonumber \\
&= P_{64} \cap \bar{P}_1 \cap P_1 \cap P_2 \cap \bar{P}_2 , \\[6pt]
I_{17} &= \Big\langle \langle 1|2+3|1],  \langle 2|3|5+6|4|2]-\langle 3|1|5+6|4|3]  \Big\rangle_{R_6} \nonumber \\
&= P_{19}(412356) \cap \bar{P}_8(321456) \cap P_8(321456) \cap \bar{P}_1 \cap P_1 , \\[6pt]
I_{18} &= \Big\langle \langle 1|3+4|2],  \langle 2|3|5+6|1|2]-\langle 3|4|5+6|1|3]  \Big\rangle_{R_6} \\
&= P_{65} \cap \bar{P}_8(561234) \cap P_1(156234) \cap P_2 \cap \bar{P}_1(234156) , \\[6pt]
I_{19} &= \Big\langle \langle 1|3+4|2],  \langle 3|2|5+6|1|3]-\langle 2|4|5+6|1|2]  \Big\rangle_{R_6} \nonumber \\
&= P_{11}(241356) \cap \bar{P}_8(561234) \cap \bar{P}_7 \cap P_6(132456) \\
&\quad \cap \bar{P}_2(213456) \cap P_2 \cap \bar{P}_1(234156) \cap P_1(156234) , \\[6pt]
I_{20} &= \Big\langle \langle 1|3+4|2],  \langle 3|4|5+6|1|3]-\langle 4|2|5+6|1|4]  \Big\rangle_{R_6} \nonumber \\
&= \bar{P}_{11}(132456) \cap \bar{P}_8(561234) \cap P_8(432165) \cap P_6(132456) \\
&\quad \cap P_2 \cap P_1(156234) \cap \bar{P}_1(234156) \cap P_1(134256) , \\[6pt]
I_{21} &= \Big\langle \langle 1|3+4|2],  \langle 1|2|5+6|3|1]-\langle 2|4|5+6|3|2]  \Big\rangle_{R_6} \nonumber \\
&= P_{66} \cap P_6(132456) \cap \bar{P}_2(213456) , \\[6pt]
I_{22} &= \Big\langle \langle 1|3+4|2],  \langle 1|4|5+6|3|1]-\langle 4|2|5+6|3|4]  \Big\rangle_{R_6} \nonumber \\
&= P_{67} \cap P_1(134256) \cap P_6(132456) , \\[6pt]
I_{23} &= \Big\langle \langle 1|3+4|2],  \langle 4|1|5+6|3|4]-\langle 1|2|5+6|3|1]  \Big\rangle_{R_6} \nonumber \\
&= P_{68} \cap P_2 \cap P_1(134256) , \\[6pt]
I_{24} &= \Big\langle s_{123},  \langle 2|3|5+6|1|2]-\langle 3|4|5+6|1|3]  \Big\rangle_{R_6} \nonumber \\
&= P_{69} \cap P_1 \cap \bar{P}_1 , \\[6pt]
I_{25} &= \Big\langle s_{123},  \langle 2|4|5+6|1|2]-\langle 4|3|5+6|1|4]  \Big\rangle_{R_6} \nonumber \\
&= P_{72} , \\[6pt]
I_{26} &= \Big\langle s_{123},  \langle 4|2|5+6|1|4]-\langle 2|3|5+6|1|2]  \Big\rangle_{R_6} \nonumber \\
&= P_{73}, \\[6pt]
I_{27} &= \Big\langle s_{123},  \langle 1|2|5+6|4|1]-\langle 2|3|5+6|4|2]  \Big\rangle_{R_6} \nonumber \\
&= \bar{P}_8(654321) \cap \bar{P}_8 \cap P_8(654321) \cap P_8 \\
&\quad \cap \bar{P}_1(456123) \cap \bar{P}_1 \cap P_1(456123) \cap P_1 .
\end{align}
where $P_1$, $P_2$, $P_5$, $P_6$, $P_7$, $P_8$, $P_{10}$, $P_{11}$, $P_{19}$ are from \cite{DeLaurentis:2025dxw} and are defined as

\begin{align}
P_1 &= \big\langle \langle 12\rangle, \langle 13\rangle, \langle 23\rangle \big\rangle_{R_6} \,, \label{eq:P1}\\
P_2 &= \big\langle \lambda_1^\alpha \big\rangle_{R_6} = \big\langle |1\rangle \big\rangle_{R_6} \,, \label{eq:P2}\\
P_5 &= \big\langle \langle 12\rangle, [13] \big\rangle_{R_6} \,, \label{eq:P5}\\
P_6 &= \big\langle \langle 12\rangle, [34] \big\rangle_{R_6} \,, \label{eq:P6}\\
P_7 &= \big\langle \langle 12\rangle, \langle 1|3+4|1], \langle 2|3+4|1] \big\rangle_{R_6} \,, \label{eq:P7}\\
P_8 &= \big\langle \langle 12\rangle, \lambda_1^\alpha [13] + \lambda_2^\alpha [23] \big\rangle_{R_6} \,, \label{eq:P8}\\
P_{10} &= \big\langle \langle 12\rangle, \langle 3|4+5|3] \big\rangle_{R_6} \,, \label{eq:P10}\\
P_{11} &= \big\langle \langle 12\rangle, \langle 3|2+4|1] \big\rangle_{R_6} \,, \label{eq:P11}\\
P_{19} &= \big\langle \langle 1|2+3|1], \langle 2|3+4|2] \big\rangle_{R_6} \,. \label{eq:P19}
\end{align}
while the newly identified primes are given by

\begin{align*}
P_{57} &= \Big\langle \langle 1|2\rangle, \\
&\quad -\langle 2|3\rangle\langle 2|4\rangle[1|4][2|3]-\langle 2|4\rangle^2[1|4][2|4]+\langle 2|3\rangle\langle 3|4\rangle[1|3][3|4], \\
&\quad -\langle 1|3\rangle\langle 2|4\rangle[1|4][2|3]-\langle 1|4\rangle\langle 2|4\rangle[1|4][2|4]+\langle 1|3\rangle\langle 3|4\rangle[1|3][3|4] \Big\rangle_{R_6}
\end{align*}

\begin{align*}
P_{58} &= \Big\langle \langle 1|2\rangle, \quad |1\rangle[1|4]\langle 3|1+2|3]+|2\rangle[2|4]\langle 3|4\rangle[3|4] \Big\rangle_{R_6}
\end{align*}

\begin{align*}
P_{59} &= \Big\langle \langle 1|2+3|1], \quad \langle 2|3|5+6|1|2]-\langle 3|4|5+6|1|3], \\
&\quad -\langle 1|2\rangle\langle 1|3\rangle\langle 2|3\rangle[2|3]+\langle 1|3\rangle^2\langle 2|4\rangle[3|4]-\langle 1|2\rangle^2\langle 3|4\rangle[2|4]-2\langle 1|2\rangle\langle 1|3\rangle\langle 3|4\rangle[3|4], \\
&\quad -\langle 2|3\rangle[1|2][1|3][2|3]-\langle 2|4\rangle[1|2]^2[3|4]+\langle 3|4\rangle[1|3]^2[2|4]-2\langle 3|4\rangle[1|2][1|3][3|4] \Big\rangle_{R_6}
\end{align*}

\begin{align*}
P_{60} &= \Big\langle \langle 1|2+3|1], \quad \langle 4|2|5+6|1|4]-\langle 2|3|5+6|1|2], \\
&\quad \langle 1|2\rangle\langle 1|3\rangle[2|3]s_{234} + \langle 1|4\rangle\langle 1|2+3|4]s_{24}, \\
&\quad [1|2][1|3]\langle 2|3\rangle s_{234} + [1|4][1|2+3|4\rangle s_{24} \Big\rangle_{R_6}
\end{align*}

\begin{align*}
P_{61} = \Big\langle& \langle 1|2+3|1], \quad \langle 1|3|5+6|2|1]-\langle 3|4|5+6|2|3], \\
&\; |1]⟨1|2⟩²⟨3|1+4|2]+⟨2|3⟩[3|4]\big(|1]⟨1|3⟩⟨1|4⟩-|2+3|1⟩⟨3|4⟩\big), \quad \lambda_i^\alpha \leftrightarrow \tilde\lambda_i^{\dot\alpha} \Big\rangle_{R_6}
\end{align*}

\begin{align*}
P_{62} &= \Big\langle \langle 1|2+3|1], \quad \langle 4|1|5+6|2|4]-\langle 1|3|5+6|2|1], \\
&\quad (s_{24}-s_{12})\langle 1|2\rangle\langle 1|3\rangle[2|3]+s_{24}\langle 1|4\rangle\langle 1|2+3|4]+[2|3](\langle 1|2\rangle^2\langle 3|4\rangle[2|4]-\langle 1|3\rangle^2[3|4]\langle 2|4\rangle), \\
&\quad (s_{24}-s_{12})[1|2][1|3]\langle 2|3\rangle+s_{24}[1|4][1|2+3|4\rangle+\langle 2|3\rangle([1|2]^2[3|4]\langle 2|4\rangle-[1|3]^2\langle 3|4\rangle[2|4]) \Big\rangle_{R_6}
\end{align*}

\begin{align*}
P_{63} &= \Big\langle \langle 1|2+3|1], \quad \langle 1|2|5+6|4|1]-\langle 2|3|5+6|4|2], \\
&\quad |1]\langle 1|2\rangle\langle 1|3\rangle\langle 4|5+6|4]+s_{24}\langle 2|3\rangle(|2+3|1\rangle), \\
&\quad |1\rangle[1|2][1|3]\langle 4|5+6|4]+s_{24}[2|3](|2+3|1]) \Big\rangle_{R_6}
\end{align*}

\begin{align*}
P_{64} &= \Big\langle \langle 1|2+3|1], \quad \langle 2|1|5+6|4|2]-\langle 1|3|5+6|4|1], \\
&\quad \langle 1|2\rangle\langle 3|1\rangle(s_{14}+2s_{24})+\langle 1|3\rangle^2\langle 2|4\rangle[3|4]-\langle 1|2\rangle^2\langle 3|4\rangle[2|4], \\
&\quad [1|2][3|1](s_{14}+2s_{24})+[1|3]^2[2|4]\langle 3|4\rangle-[1|2]^2[3|4]\langle 2|4\rangle \Big\rangle_{R_6}
\end{align*}

\begin{align*}
P_{65} &= \Big\langle \langle 1|3+4|2], \quad \langle 1|3\rangle[1|2+3|4\rangle-\langle 1|2\rangle\langle 3|4\rangle[1|2], \quad s_{23}[1|2]+\langle 3|4\rangle[1|3][2|4] \Big\rangle_{R_6}
\end{align*}

\begin{align*}
P_{66} = \Big\langle & ⟨1|3+4|2], \; ⟨1|2|5+6|3|1]-⟨2|4|5+6|3|2], \\
&\quad ⟨4|1|2]⟨2|1+4|3]+s_{23}⟨4|1+2|3], \; ⟨1|2⟩[1|3]s_{134}+⟨2|4⟩[3|4](s_{14}-s_{23}) \Big\rangle_{R_6}
\end{align*}

\begin{align*}
P_{67} &= \Big\langle \langle 1|3+4|2], \quad |5+6|2]⟨2|4⟩[3|4]+|1|3](s_{23}-s_{14}) \Big\rangle_{R_6}
\end{align*}

\begin{align*}
P_{68} &= \Big\langle \langle 1|3+4|2], \quad \langle 4|1|5+6|3|4]-\langle 1|2|5+6|3|1], \\
&\quad [1|3][2|4]\langle 2|1+4|2]+[1|4][2|3]\langle 3|2+4|3] \Big\rangle_{R_6}
\end{align*}

\begin{align*}
P_{69} &= \Big\langle s_{123}, \quad \langle 2|3|5+6|1|2]-\langle 3|4|5+6|1|3], \\
&\quad [1|(\langle 2|1\rangle\langle 1|3\rangle\langle 3|2+4|3]+[4|3]\langle 3|1\rangle(\langle 1|3\rangle\langle 4|2\rangle+\langle 2|1\rangle\langle 4|3\rangle)+[4|2]\langle 2|1\rangle\langle 2|1\rangle\langle 4|3\rangle) \\
&\quad +[2|([4|3]\langle 1|3\rangle\langle 3|2\rangle\langle 4|2\rangle + \langle 2|1\rangle\langle 2|3\rangle(s_{34}+s_{23}))+[3|\langle 3|2\rangle\langle 3|4\rangle\langle 1|2+3|4], \\
&\quad |1\rangle([1|2][3|1]\langle 3|2+4|3]+\langle 3|4\rangle[1|3]([3|1][2|4]+[1|2][3|4])+\langle 2|4\rangle[1|2][1|2][3|4]) \\
&\quad +|2\rangle(\langle 3|4\rangle[3|1][2|3][2|4] + [1|2][3|2](s_{34}+s_{23}))+|3\rangle[2|3][4|3][1|2+3|4\rangle \Big\rangle_{R_6}
\end{align*}
\end{appendix}
\bibliographystyle{JHEP}
\bibliography{main.bib}

\end{document}